\documentclass{emulateapj}
\usepackage{aas_macros}
\usepackage{apjfonts}
\usepackage{html}
\usepackage{graphicx}
\usepackage[squaren]{SIunits}

\submitted{Accepted for publication in ApJ}

\def\bq{\begin{equation}} 
\def\eq{\end{equation}}
\newcommand{\bqa}{\begin{eqnarray}} 
\newcommand{\eqa}{\end{eqnarray}}
\newcommand{\nn}{\nonumber \\}
\newcommand{\hacc}{{\small HACC}}
\newcommand{\crkhacc}{{\small CRK-HACC}}
\newcommand{\gadget}{{\small GADGET}}
\newcommand{\asura}{{\small ASURA}}
\newcommand{\gasoline}{{\small GASOLINE}}
\newcommand{\ramses}{{\small RAMSES}}
\newcommand{\enzo}{{\small ENZO}}
\newcommand{\nyx}{{Nyx}}
\newcommand{\arepo}{{\small AREPO}}
\newcommand{\gizmo}{{\small GIZMO}}
\newcommand{\art}{{\small ART}}
\newcommand{\camb}{{\small CAMB}}
\newcommand{\healpix}{{\small HEALPIX}}

\newcommand{\planck}{{\em Planck}}

\addunit{\massh}{\mathit{h}^{-1}M_\odot}
\addunit{\Mpch}{\mathit{h}^{-1}Mpc}
\addunit{\kpch}{\mathit{h}^{-1}kpc}
\addunit{\kms}{km\ s^{-1}}
\addunit{\invMpch}{\mathit{h}\ Mpc^{-1}}
\addunit{\GHz}{GHz}
\addunit{\sqamin}{arcmin^2}
\addunit{\sqdeg}{deg^2}

\def\smhwidth{0.45\textwidth}
\def\smfwidth{0.90\textwidth}

\shorttitle{The Borg Cube Simulation}

\begin{document}

\title{The Borg Cube Simulation: Cosmological Hydrodynamics with CRK-SPH}

\author{J.D.~Emberson\altaffilmark{1,2}, Nicholas~Frontiere\altaffilmark{2,3}, 
Salman~Habib\altaffilmark{2,4}, Katrin~Heitmann\altaffilmark{2,4}, Patricia~Larsen\altaffilmark{2}, 
Hal~Finkel\altaffilmark{1}, Adrian~Pope\altaffilmark{5}}
\altaffiltext{1}{ALCF Division, Argonne National Laboratory, Lemont, IL 60439, USA}
\altaffiltext{2}{HEP Division, Argonne National Laboratory, Lemont, IL 60439, USA}
\altaffiltext{3}{Department of Physics, University of Chicago, Chicago, IL 60637, USA} 
\altaffiltext{4}{MCS Division, Argonne National Laboratory, Lemont, IL 60439, USA}
\altaffiltext{5}{CPS Division, Argonne National Laboratory, Lemont, IL 60439, USA}
\email{jemberson@anl.gov}

\begin{abstract}
A challenging requirement posed by next-generation observations is a firm
theoretical grasp of the impact of baryons on structure formation.
Cosmological hydrodynamic simulations modeling gas physics
are vital in this regard.
A high degree of modeling flexibility exists in this space making it important
to explore a range of methods in order to gauge the 
accuracy of simulation predictions.
We present results from the first cosmological simulation using 
Conservative Reproducing Kernel Smoothed Particle Hydrodynamics (CRK-SPH). 
We employ two simulations: one evolved purely
under gravity and the other with non-radiative hydrodynamics. Each
contains $2\times2304^3$ cold dark matter plus baryon particles in an $\unit{800}{\Mpch}$
box. We compare statistics to previous non-radiative simulations including power spectra,
mass functions, baryon fractions, and concentration. We find self-similar radial
profiles of gas temperature, entropy, and pressure and show that a simple analytic 
model recovers these results to better than 
$40\%$ over two orders of magnitude in mass.
We quantify the level of 
non-thermal pressure support in halos and demonstrate that hydrostatic mass 
estimates are biased low by $24\%$ ($10\%$) for halos of mass 
$\unit{10^{15}~(10^{13})}{\massh}$. We compute angular power spectra for the thermal
and kinematic Sunyaev-Zel'dovich effects and find good agreement with the low-$\ell$
\planck\ measurements. Finally, artificial scattering between
particles of unequal mass is shown to have a large impact on the gravity-only 
run and we highlight the importance of better understanding this issue in 
hydrodynamic applications.
This is the first in a simulation campaign using CRK-SPH with
future work including subresolution gas treatments.
\end{abstract}



\section{Introduction}

It has been twenty years since the discovery of the accelerated expansion of 
the universe driven by dark energy \citep{riess/etal:1998,perlmutter/etal:1999}. 
Today, the nature of dark energy remains unknown on a fundamental level despite
the fact that it presently dominates the Universe's energy budget at a level of 
$\approx70\%$ \citep[e.g.,][]{planck/etal:2013a}. This puzzle has motivated a concerted effort 
to investigate the nature of dark energy through its influence on the expansion 
and structure growth histories of the universe
\citep[see e.g.,][for a recent review]{weinberg/etal:2013} enabled by current and upcoming sky surveys that provide large-scale
structure statistics at unprecedented levels of detail and statistical precision. These campaigns include
BOSS \citep[Baryon Oscillation Spectroscopic Survey;][]{boss/etal:2013}, 
DES \citep[Dark Energy Survey;][]{des/etal:2005},
DESI \citep[Dark Energy Spectroscopic Instrument;][]{levi/etal:2013},
LSST \citep[Large Synoptic Survey Telescope;][]{lsst/etal:2009}, 
Euclid \citep{euclid/etal:2011}, 
and WFIRST \citep[Wide-Field InfraRed Survey Telescope;][]{wfirst/etal:2013}.
Given the very low level of statistical uncertainty in the observations, theoretical and modeling systematics are a significant source of concern -- as are measurement systematics -- when considering the ultimate limits on the scientific information that can be gleaned from the surveys. It is therefore necessary that theory, modeling, and simulations be developed to a new level of detail and robustness in order to meet the varied challenges posed by next-generation observations.

The corresponding computational challenge is formidable due to the extreme range of spatio-temporal
scales over which accurate predictions are needed. Generally speaking, we need
to model volumes comparable to survey depths with sufficient resolution
to capture the scales relevant for galaxy formation. The essential tools for this task 
are cosmological simulations that probe the deeply nonlinear regime of structure 
formation. The simplest examples are $N$-body simulations where 
the dynamics are dictated solely by gravitational forces. These simulations have
become ubiquitous in the field and have proven useful in a wide variety of 
applications. The common convention assumes a potential dominated by cold
dark matter (CDM) though other models inducing large changes on small scales
such as warm dark matter or modified gravity scenarios are also studied. In any 
case, one certainly important omission in gravity-only (GO) simulations are contributions 
from baryons. In contrast to dark matter, baryons can dissipate energy, 
allowing them to cool and condense on small scales. The ensuing star formation and 
feedback from supernova and active galactic
nuclei (AGN) have the potential to alter the matter distribution on small to 
moderately large scales. Separating the effects of complex astrophysical mechanisms from the 
fundamental physics associated with dark energy (and potentially dark matter physics at small scales) is essential for the proper interpretation of 
observations.

To this end, numerous cosmology codes are equipped with hydrodynamic solvers 
to model baryonic physics. Typically, this is done using either particle-based
Lagrangian methods with smoothed particle hydrodynamics (SPH)
(e.g., \gadget; \citealt{springel:2005}, \asura; \citealt{saitoh/etal:2008},
\gasoline; \citealt{wadsley/etal:2017}),
or mesh-based Eulerian methods on a stationary grid with possible adaptive 
mesh refinement (AMR) (e.g., \art; \citealt{kravtsov/etal:1997},
\ramses; \citealt{teyssier:2002}, \nyx; \citealt{almgren/etal:2013},
\enzo; \citealt{bryan/etal:2014}).
Hybrid schemes incorporating features of both SPH and AMR also
exist (\arepo; \citealt{springel:2010}, \gizmo; \citealt{hopkins:2015}).

The most basic use case is that of non-radiative (NR) or ``adiabatic'' hydrodynamics 
in which the thermal state of baryons changes only in response to gravitational shock 
heating and adiabatic cooling with the expansion of the universe. 
While the pure NR case is a simplified description of the real universe that omits a number
of important physics, it is still accurate on large to quasi-small length scales 
($k \lesssim \unit{1}{\invMpch}$).
For instance, \citet{burns/etal:2010} showed that NR simulations are sufficient to match 
the thermodynamic properties of gas on the outskirts of observed clusters. 
Moreover, NR physics is the natural first step of a hydrodynamics solver since this
represents the most parameter-free reference point for comparison with other codes.
Unfortunately, even in this most 
simple case, the Santa Barbara code comparison project \citep{frenk/etal:1999}
showed that systematic differences arise between Eulerian and Lagrangian 
methods when simulating a massive galaxy cluster. The more recent nIFTy code
comparison project \citep{sembolini/etal:2016} showed that much of this
discrepancy has been resolved with modern SPH treatments matching
more closely with mesh codes, though small differences
still exist. Code comparisons like these are important in quantifying the level
of confidence in simulation results. This becomes increasingly true as more
complicated physics such as cooling and feedback are added, since 
these treatments vary substantially amongst codes. 

It is therefore important to proceed in a controlled manner when designing a new
cosmological hydrodynamics code. This paper is the first in a series to
expand the gravitational framework of the Hardware/Hybrid Accelerated Cosmology Code
\citep[\hacc;][]{habib/etal:2016} with a hydro solver equipped with a full suite of 
subresolution gas physics models capable of scaling to the problem size
demanded by upcoming observations. Here we begin with an exclusive focus 
on the NR case, which serves as an important first step in the systematic process of 
modeling all of the baryonic physics relevant on cosmological scales. We perform 
both NR and GO simulations in large volumes with high mass resolution and compare 
to previous NR simulations to evaluate where we stand in relation to other methods.
Furthermore, the large simulation volume used here allows us to significantly expand the 
statistical analysis of group and cluster-scale halos compared to previous NR runs that used 
smaller boxes. Similarly, the large box resolves additional power from low-$k$ density and
velocity modes, enabling the construction of more accurate synthetic sky maps such as
the thermal and kinematic components of the Sunyaev-Zel'dovich effect.

We employ a hydro solver based on the Conservative Reproducing Kernel Smoothed Particle
Hydrodynamics \citep[CRK-SPH;][]{frontiere/etal:2017} algorithm. CRK-SPH is a modern SPH 
treatment that overcomes two of the main shortcomings of traditional SPH (tSPH); 
namely, zeroth-order inaccuracy and an overly aggressive artificial viscosity. 
\citet{frontiere/etal:2017} demonstrate this method to work robustly on a wide range of
hydrodynamic tests including those with strong shocks and dynamical fluid instabilities.
In addition, \cite{raskin/owen:2016} recently showed it to outperform standard SPH 
treatments in the astrophysically relevant case of a generalized 
rotating disk. 
Here we present results from the first use of CRK-SPH in a large-scale 
cosmological setting and show that it agrees well with other modern methods.

The incorporation of CRK-SPH into {\hacc} has been designed from the 
outset for high performance and full scalability on next-generation supercomputers 
and is capable of running on all current high performance computing architectures.
We refer to the new gravity plus hydro framework as {\crkhacc} and provide technical
details of its implementation as well as results from test cases such as the Santa
Barbara run in \citet{frontiere/etal:inprep}.

This paper is organized as follows. In Section~\ref{sec:method}, we provide details
of the simulation setup and an overview of our numerical methods. 
Section~\ref{sec:globstats} presents summary statistics from the main run with 
an emphasis on comparisons between the NR and GO cases. These include power spectra, 
halo mass functions, baryon fractions, and halo concentrations. 
Section~\ref{sec:gprof} examines the gaseous components of halos and shows that
a simple analytic model is able to match the simulated
density, temperature, entropy, and pressure profiles with reasonably high accuracy.
Section~\ref{sec:hse} measures the hydrostatic mass bias and fraction of non-thermal
pressure support for group and cluster-scale halos.
Section~\ref{sec:sz} provides an analysis of the thermal and kinematic components of 
the Sunyaev-Zel'dovich effect. 
We finish with a summary and conclusions of our work in Section~\ref{sec:summary}.


\section{Numerical Method}
\label{sec:method}

We perform two simulations utilizing the same initial conditions in order
to compare the GO and NR cases
of structure formation. Both simulations evolve $2304^3$ each of CDM
and baryon particles in a box of side length $\unit{800}{\Mpch}$
from redshift $z = 200$ to 0. Each species is initialized on a uniform
mesh and displaced using the Zel'dovich approximation \citep{zeldovich:1970} 
with a random realization drawn from the combined CDM plus baryon transfer function
generated using \camb\ \citep{lewis/etal:2000}. The initial CDM and baryon meshes 
are maximally offset by staggering them by half the mean interparticle
separation in each dimension. This is done to minimize artificial particle
coupling in the initial conditions \citep{yoshida/etal:2003}. Both meshes
draw from the same white noise field and we account for the phase shift in the 
staggered grid when assigning displacements and velocities \citep{valkenburg/etal:2017}. 
Finally, we set the initial thermal energy of baryons in the NR run to a uniform 
temperature of $\unit{847}{\kelvin}$ so as to match the adiabatic relation of gas 
that follows the cosmic microwave background (CMB) temperature until decoupling
and adiabatically cooling at $z = 129$. 

The GO simulation is performed using \hacc\ as a standard $N$-body run with
dynamics based solely on gravity. The NR simulation is run with \crkhacc\ 
and subjects baryons to both gravitational and hydrodynamic forces.
We refer to the pair of runs as the Borg Cube simulations. 
Henceforth, quantities associated with the GO run are
appended with a ``go'' subscript while quantities without subscripts refer
to the NR case. Throughout our work, we employ the best-fit WMAP-7 cosmology
\citep{komatsu/etal:2011} with ($\Omega_c$, $\Omega_b$, $\Omega_\Lambda$, $\Omega_\nu$,
$w$, $n_s$, $\sigma_8$, $h$) = (0.22, 0.0448, 0.7352, 0, -1, 0.963, 0.8, 0.71),
which was previously used in the Q Continuum \citep{heitmann/etal:2015}, 
Outer Rim \citep{habib/etal:2016}, and Mira Universe \citep{heitmann/etal:2016}
simulations run with \hacc. Particle masses in each simulation are
$m_{c} = \unit{2.56\times10^9}{\massh}$ and $m_{b} = \unit{5.21\times10^8}{\massh}$
for CDM and baryons, respectively. This allows for individual group and cluster-scale 
halos (i.e., masses $\gtrsim \unit{10^{13}}{\massh}$) to be resolved with at least 
thousands of particles. Moreover, with 138,000 such halos at $z = 0$, we are able to
stack halo profiles in multiple mass bins to obtain much more statistically robust results 
compared to previous simulations run with smaller simulation volumes.
The gas in the NR run is modeled using an adiabatic index $\gamma = 5/3$ with 
all CRK-SPH parameters (Courant factor, viscosity coefficients, etc.) 
following directly from \citet{frontiere/etal:2017}.

The gravitational force resolution is determined by the softening length. For 
multi-species runs, one must be careful to find a balance in force 
softening that mitigates artificial coupling between particles of unequal mass
\citep{angulo/etal:2013} while maintaining adequate resolution.
On the one hand, too small softening will cause low-mass particles to
scatter off their high-mass counterparts. On the other hand, too large softening will
inhibit structure growth on small scales due to the oversmoothing. 
For the Borg Cube runs, we use a Plummer softening length of
$r_{\rm soft} = \unit{14}{\kpch}$ (in comoving units), which
is $1/25$ of the mean interparticle separation. 
This is held constant in time and between all particle pairs. 
One alternative approach involves using the SPH smoothing length to
set a spatially adaptive softening length. The downside of this approach
is that CDM-baryon gravitational interactions are heavily suppressed at early 
times when the particle distribution is relatively homogeneous
so that the smoothing length is comparable to the mean
interparticle separation \citep{angulo/etal:2013,villaescusa/etal:2017}.

\begin{figure}
\begin{center}
\includegraphics[width=\smhwidth]{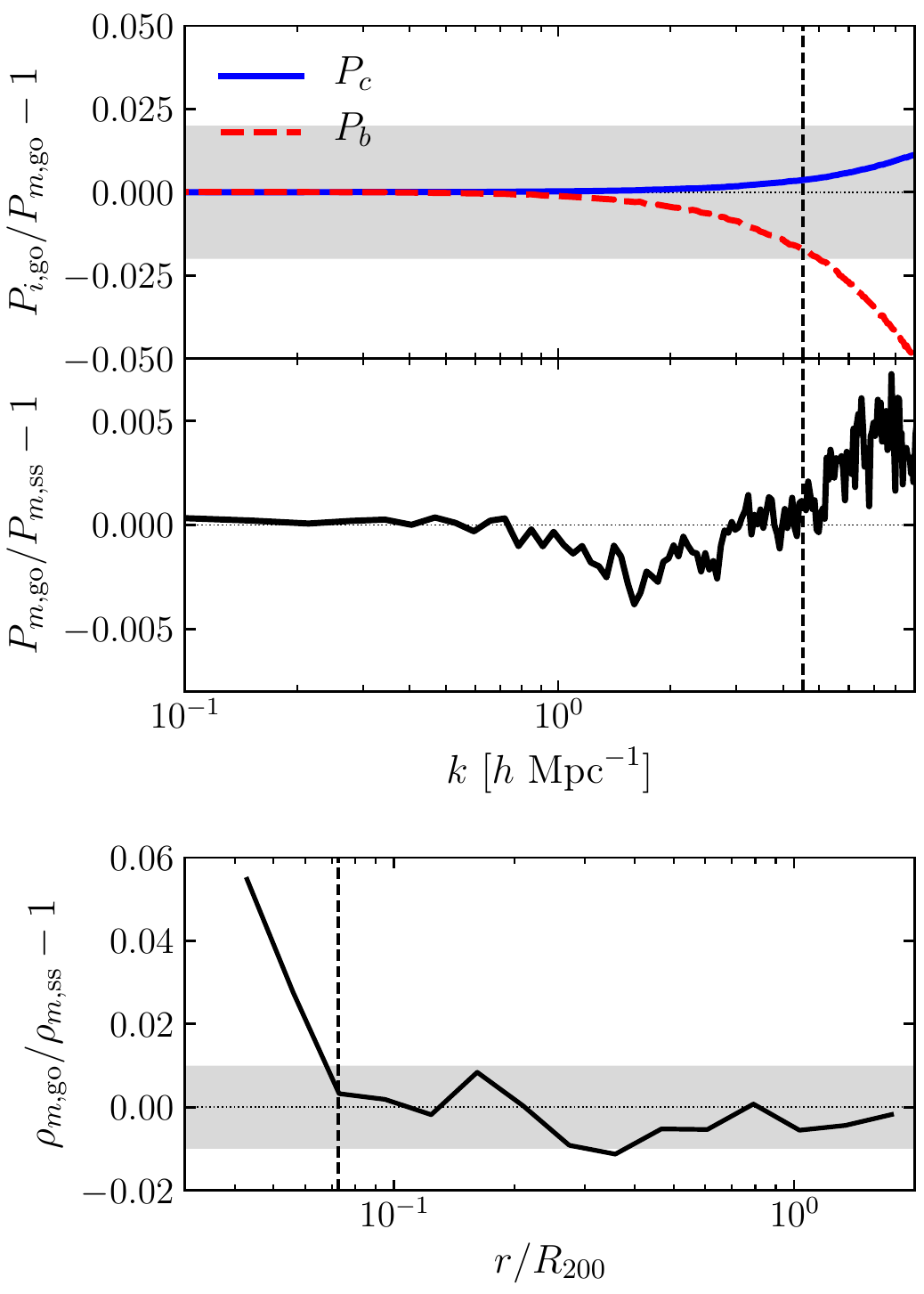}
\end{center}
\caption{Top panel shows the difference in power for the CDM (solid blue) and baryon 
(dashed red) components of the down-scaled GO run relative to the total matter at $z = 0$. 
Deviations are within $\pm2\%$ up to half the particle Nyquist frequency which is 
denoted by the vertical dashed black line. The peeling away of the the two components 
for $k \gtrsim \unit{1}{\invMpch}$ is caused by artificial scattering 
between the two species. This run used a constant $r_{\rm soft} = \unit{14}{\kpch}$;
deviations become stronger and shifted to larger scales for finer softening.
Middle panel compares the total matter from the down-scaled GO and SS runs 
at $z = 0$. Although the individual components of the GO run experience percent-scale
deviations, the total matter remains relatively unchanged compared to the SS case.
Bottom panel compares stacked radial profiles of the total matter density for all halos
with masses between $\unit{10^{13}-10^{13.5}}{\massh}$ for the GO and SS
runs at $z = 0$. The profiles are truncated below $r_{\rm soft}$ and the vertical 
dashed line denotes the scale $\unit{30}{\kpch}$ above which the profiles agree to
roughly $1\%$.}
\label{fig:powerrsoft}
\end{figure}

Our choice of $r_{\rm soft}$ was made by finding the smallest value in the GO
run such that differences in the CDM and baryon power relative to the total matter 
are within $2\%$ up to half the particle Nyquist frequency. This was achieved by 
experimenting on a down-scaled version of the Borg Cube with $2\times288^3$ 
particles in a box of side length $\unit{100}{\Mpch}$. The top panel of Figure 
\ref{fig:powerrsoft} shows the relative difference in power for each component 
in the fiducial case where $r_{\rm soft} = \unit{14}{\kpch}$.
Ideally, we expect the two species to have identical power since they were 
initialized with the same transfer function and hydro forces are absent. 
Instead, scattering between heavy CDM and light baryons induces an enhancement 
(suppression) in power for the former (latter) on small scales. Using a smaller 
softening amplifies these deviations and shifts the onset of the problem to larger
scales.

Interestingly, this artificial mass segregation operates in such a way that the total matter
distribution is relatively unchanged. To show this, we ran a single-species (SS)
simulation with $1\times288^3$ particles representing the combined CDM plus baryon 
field (i.e., the setup of a traditional $N$-body simulation). The middle panel of 
Figure \ref{fig:powerrsoft} compares the total matter power from the GO and SS
runs; differences are sub-percent on all scales. \citet{angulo/etal:2013} note a
similar finding in their investigation of particle coupling in GO simulations.
This invariance also emerges when examining radial profiles of the total matter content
within halos. The bottom panel of Figure \ref{fig:powerrsoft} shows the relative difference 
between GO and SS density profiles stacked over all halos in the mass range
$\unit{10^{13}-10^{13.5}}{\massh}$ (208 halos in GO; 206 in SS).
These agree to about 
$1\%$ down to $r = \unit{30}{\kpch} \approx 2r_{\rm soft}$. On smaller scales, 
the GO simulation displays systematically higher density at a level approaching 
$6\%$ at $r_{\rm soft}$.

We reiterate that this analysis was focused on suppressing the effects of mass 
segregation resulting from artificially strong interactions between unequal mass particles. 
In principle, other discreteness effects (e.g., strong interactions between particles of equal
mass) that would also impact ordinary SS $N$-body simulations could also be present here.
Such discreteness effects are an important topic 
\citep[see, e.g.,][]{melott:2007,power/etal:2016}
and we leave to future work a more thorough 
investigation of their impact specifically in regard to hydrodynamic simulations.

Another important aspect to consider is the relationship between gravitational
and hydrodynamic resolutions. As mentioned earlier, we use a constant softening
length for all particle pairs, including both CDM and baryons. This means that
there is a mismatch between the constant gravitational force resolution 
and the changing hydrodynamic resolution set by the adaptive SPH 
smoothing lengths, $h$. In this case, one must be careful to avoid the situation
where $h$ drops below $r_{\rm soft}$ since this results in an unphysical numerical
setup with hydrodynamics being evolved below the resolution limit of gravity.
The alternate case where $h > r_{\rm soft}$ can also lead to unphysical
fragmentation of gas clouds in the event that $h$ is close to the Jeans length 
\citep{bate/burkert:1997}. 
Both of these scenarios are avoided here since the minimum smoothing length measured 
in the Borg Cube is $h_{\rm min} = \unit{26}{\kpch} \approx 2r_{\rm soft}$ 
and the cosmological Jeans length is much smaller than the scales resolved here. 
Of course, it will be important to revisit this topic in future work with
additional physics since cooling will significantly reduce $h_{\rm min}$ and other 
conditions -- such as ensuring that the critical density for star formation is 
correctly resolved \citep[e.g.,][]{hopkins/etal:2018} -- must be met.

Much of our analysis requires the identification of halos. This is achieved in
a two-step fashion using the CosmoTools parallel analysis framework within \hacc. 
First, we run a friends-of-friends (FOF) finder with a linking 
length of $b = 0.168$. This is done only on the CDM particles to ensure that each
particle in the FOF group has the same mass. We designate the halo center as the
location of the most bound CDM particle in the FOF group. The next step is to create
spherical overdensity (SO) halos by starting at the location of each FOF center and
moving outwards in spherical shells until we reach the radius, $R_{200}$, at which
the interior density is 200 times the critical density of the universe. 
These SO halos are 
constructed out of both CDM and baryon tracer particles. We show later that halo properties are
converged for masses $M_{200} \geq \unit{10^{13}}{\massh}$, corresponding to
a combined mass of roughly $3200(m_c + m_b)$. A thousand-particle threshold minimum for
converged halo properties is also typical of single-species $N$-body 
simulations
\citep[e.g.,][]{power/etal:2003,child/etal:2018}.

The central density of a halo is often described in terms of its concentration.
This is defined as $c_{200} \equiv R_{\rm 200}/R_s$ where $R_s$ is the
scale radius of the SO halo. A common convention is to find the concentration
that matches the best-fit Navarro-Frenk-White \citep[NFW;][]{navarro/etal:1997}
density profile of the halo. While an NFW form is still mostly justified in the
NR case, large deviations occur once sophisticated gas treatments
such as radiative cooling and feedback are included \citep{rasia/etal:2013}.
In order to facilitate comparison with such cases, we opt for a concentration
definition that is independent of the underlying density profile. Moreover,
this allows us to compute concentrations for the baryon component of each SO
halo separately, which will deviate strongly from an NFW form even in the NR case. 

In what follows, we use the ``peak finding'' concentration method described in
\citet{child/etal:2018}. The procedure is to find the radius, $R_{\rm peak}$, 
at which the differential mass profile, $dM/dr$, is a maximum and to define 
$c_{200} \equiv R_{200}/R_{\rm peak}$. This is achieved during halo-finding
by first computing $dM/dr$ in 20 logarithmically-spaced spherical shells around
each halo. A three-point Hann filter is used to smooth the differential mass profile 
and the bin with the maximum value is identified. If this bin is an endpoint, we
set $R_{\rm peak}$ as the radius of the shell; otherwise, we fit a cubic spline
to the bin and one adjacent neighbor in each direction to numerically solve
for the peak. 

\begin{table*}[t]
\caption{Simulation Parameters for the Main Runs Referenced in this Work}
\centering
\begin{tabular}{c c c c c c c c c c c c}
\hline\hline
Simulation & Reference & Code & $\Omega_c$ & $\Omega_b$ & $\sigma_8$ & $n_s$ & $h$ & 
$L_{\rm box}$& $N_{c,b}$ & $m_c$ & $m_b$ \\
 &  &  &  &  &  &  &  & $\mathit{h}^{-1}$Mpc & & $\mathit{h}^{-1}M_\odot$ & $\mathit{h}^{-1}M_\odot$ \\ 
\hline
Borg Cube & This Paper & {\crkhacc} & 0.22 & 0.0448 & 0.8 & 0.963 & 0.71 & 800 & $2304^3$ & $2.56\times10^9$ & $5.21\times10^8$ \\
J06 & \citet{jing/etal:2006} & {\gadget} & 0.224 & 0.044 & 0.85 & 1.0 & 0.71 & 100 & $512^3$ & $4.64\times10^8$ & $9.11\times10^7$ \\
Illustris NR & \citet{vogelsberger/etal:2014} & {\arepo} & 0.227 & 0.0456 & 0.809 & 0.963 & 0.704 & 75 & $910^3$ & $3.53\times10^7$ & $7.10\times10^6$ \\
Illustris TNG & \citet{springel/etal:2018} & {\arepo} & 0.2603 & 0.0486 & 0.8159 & 0.9667 & 0.6774 & 205 & $2500^3$ & $3.99\times10^7$  & $7.45\times10^6$ \\
B10 NR & \citet{battaglia/etal:2010} & {\gadget} & 0.207 & 0.043 & 0.8 & 0.96& 0.72 & 165 & $256^3$ & $1.54\times10^{10}$ & $3.20\times10^9$ \\
B10 Feedback & \citet{battaglia/etal:2010} & {\gadget} & 0.207 & 0.043 & 0.8 & 0.96 & 0.72 & 165 & $256^3$ & $1.54\times10^{10}$ & $3.20\times10^9$ \\
\hline
\label{table:sims}
\end{tabular}
\end{table*}

In the following sections, we compare results from the Borg Cube to other cosmological hydrodynamic simulations. 
Table \ref{table:sims} provides a useful summary of the main simulations referenced here. In each case, we list the code 
used, the cosmology assumed, the box width, particle count, and mass resolution for 
both CDM and baryons. The box size is in units of $\Mpch$ while particle masses are in units of $\massh$.
The hydro solvers compared are the CRK-SPH scheme of {\crkhacc}, the tSPH implementation in {\gadget}, and the moving 
mesh method of {\arepo}. All of the simulations are run in the NR regime except for Illustris TNG and B10 Feedback which both
include additional prescriptions for radiative cooling and feedback from supernovae plus AGN. The simulations span three
orders of magnitude in volume and mass resolution with the Borg Cube having the largest volume and a mass resolution 
suitable for group and cluster-scale halos.


\section{Summary Statistics}
\label{sec:globstats}

\subsection{Power Spectra}

\begin{figure*}
\begin{center}
\includegraphics[width=0.9\textwidth]{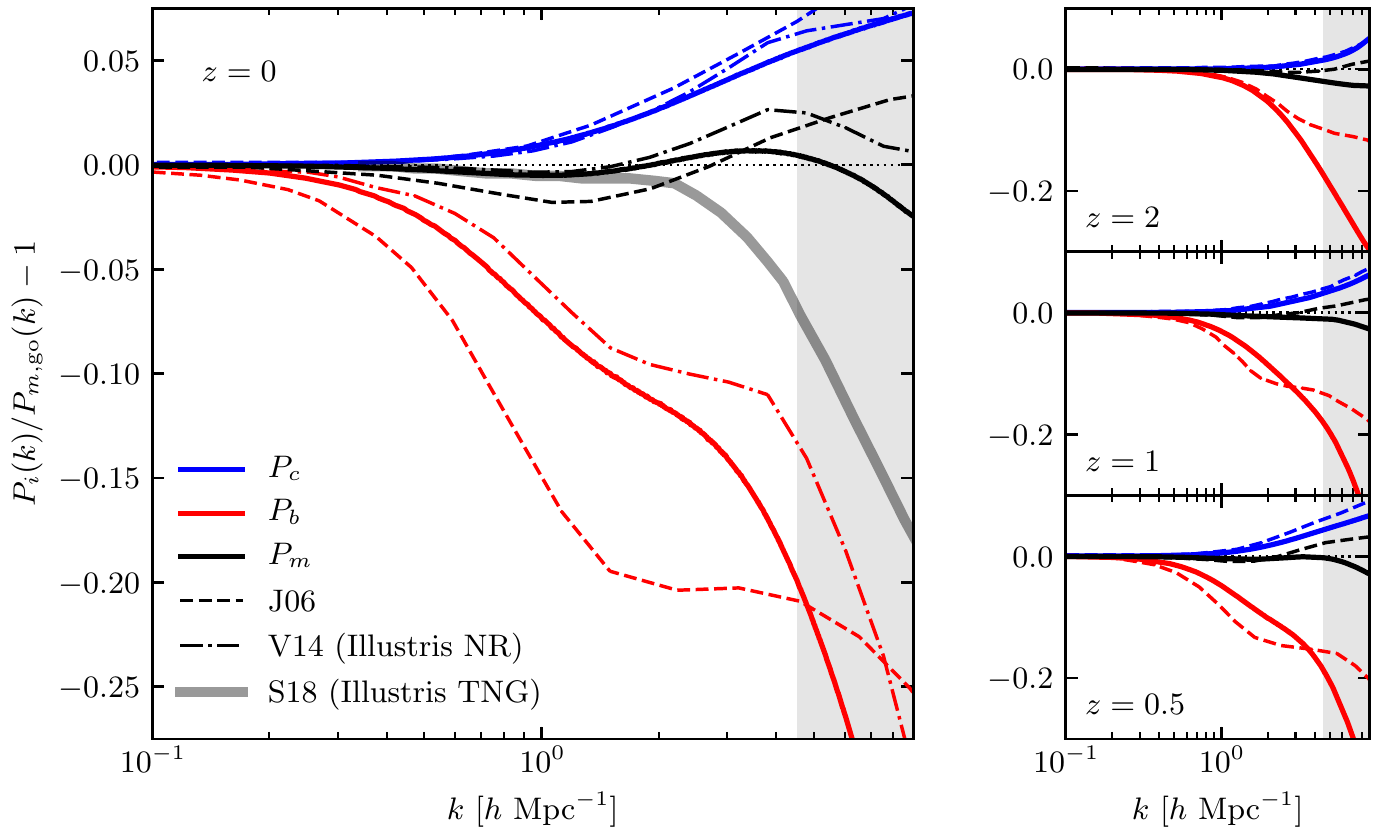}
\end{center}
\caption{Ratio in power spectra for the CDM (blue), baryon (red), and
total matter (black) components of the NR simulation relative to the total matter 
result in the GO case. Solid lines are from the Borg Cube simulation while the 
dashed and dash-dotted lines show corresponding results from the NR simulations of 
\citet{jing/etal:2006} and \citet{vogelsberger/etal:2014}, respectively. 
The thick
shaded line in the $z = 0$ panel traces the total matter result from the 
Illustris TNG cooling plus feedback simulation presented in
\citet{springel/etal:2018}. 
Note that the ratio in power for each set of simulations is measured with respect
to their own corresponding GO run.
The vertical shaded band in each panel
denotes scales above half the particle Nyquist frequency of the Borg Cube run.
}
\label{fig:bcpower}
\end{figure*}

We begin with an investigation of how NR processes affect the density power
spectrum. In Figure \ref{fig:bcpower}, we plot the ratio in power for each
component in the NR simulation relative to the total matter power in the
GO run. For comparison, we show the results from the NR simulation of
\citet[][hereafter J06]{jing/etal:2006} who performed a similar 
analysis using \gadget. The two analyses display
qualitatively similar trends. In the first place, both works find the expected 
behavior that shock-heated gas provides thermal pressure support that suppresses
baryon power on small scales as they resist gravitational collapse within the
potential wells of CDM structure. This leads to a redistribution of baryons
within collapsed objects that induces a gravitational back-reaction on the CDM
in such a way as to increase its clustering on small scales. The mechanism behind
this process is attributed to an energy exchange between the two species that
causes CDM to sink further within the potential well during halo formation
\citep{rasia/etal:2004,lin/etal:2006,mccarthy/etal:2007}.

The agreement between the Borg Cube and J06 results is most evident at early
times and on intermediate scales, as seen in the right panels of Figure 
\ref{fig:bcpower}. As time evolves, the two methods slowly depart in the details
of baryon redistribution and how this back-reacts on the CDM component.  
Most notably, the Borg Cube shows less baryon suppression on intermediate scales, 
$k \sim \unit{1}{\invMpch}$, with a steeper decline in power on smaller scales. 
Meanwhile, the enhancement in CDM power is consistently lower on all scales for 
the Borg Cube. Due to the opposing trends of baryons and CDM, the total matter
distribution is less affected by NR hydro than its constituents. In
the case of Borg Cube, the CDM and baryons conspire in such a way that the GO
and NR total matter power differ by less than $2\%$ for $k \leq \unit{4}{\invMpch}$
at all times. On smaller scales, the battle between the two components is dominated
by baryon suppression, with the total matter showing a decrease in power. The 
opposite occurs in J06 with the total matter showing an increase in power on
small scales.

Of course, we do not expect our results to exactly match those of J06
due to slight variations in cosmology, resolution and also
systematic differences in the tSPH and CRK-SPH implementations of \gadget\ 
and \crkhacc, respectively. More specifically, given that tSPH is now somewhat
outdated, we expect to find better agreement with more modern treatments. 
For example, the dash-dotted lines in Figure \ref{fig:bcpower} trace the
results from the Illustris-NR-2 simulation \citep{vogelsberger/etal:2014} evolved 
using the moving-mesh code {\arepo} (data provided courtesy of V. Springel). 
In this case, we find much better agreement
with the CDM and total matter matching extremely well for 
$k \lesssim \unit{2}{\invMpch}$. The baryon curves also agree much better though
there does appear to be a systematic difference with our result predicting
slightly more suppression at the few-percent level. 
This possibly reflects differences in the two hydro solvers though 
part of the discrepancy on small scales is also likely attributed to the coarser 
spatial resolution of Borg Cube (the Nyquist
frequency in Illustris-NR-2 is four times larger). In future work, we will 
provide results from an extensive comparison campaign with the mesh-based code 
{\nyx}, which will provide another useful reference point in gauging the relative
agreement amongst NR hydro codes.

The smallest scales in the Borg Cube run will be severely impacted by the
baryonic cooling and feedback processes it omits. On moderately
large scales, however, we expect qualitative agreement with simulations
including these contributions. For instance, the thick shaded gray line in Figure
\ref{fig:bcpower} shows the result from \citet{springel/etal:2018} for the recent
Illustris TNG300 simulation. This curve suggests that changes from additional physics
are mostly confined to scales $k \gtrsim \unit{1}{\invMpch}$ where the total matter
becomes strongly suppressed compared to the NR case\footnote{Note that
the exact scale at which the total matter power becomes suppressed is
sensitive to subresolution gas treatments, particularly in regard to the details of 
AGN feedback modeling \citep[see e.g.,][]{springel/etal:2018}.}.
This suppression results from
the depletion and redistribution of gas via star formation and feedback occurring 
on small scales.

\subsection{Halo Mass Function}

\begin{figure}
\begin{center}
\includegraphics[width=\smhwidth]{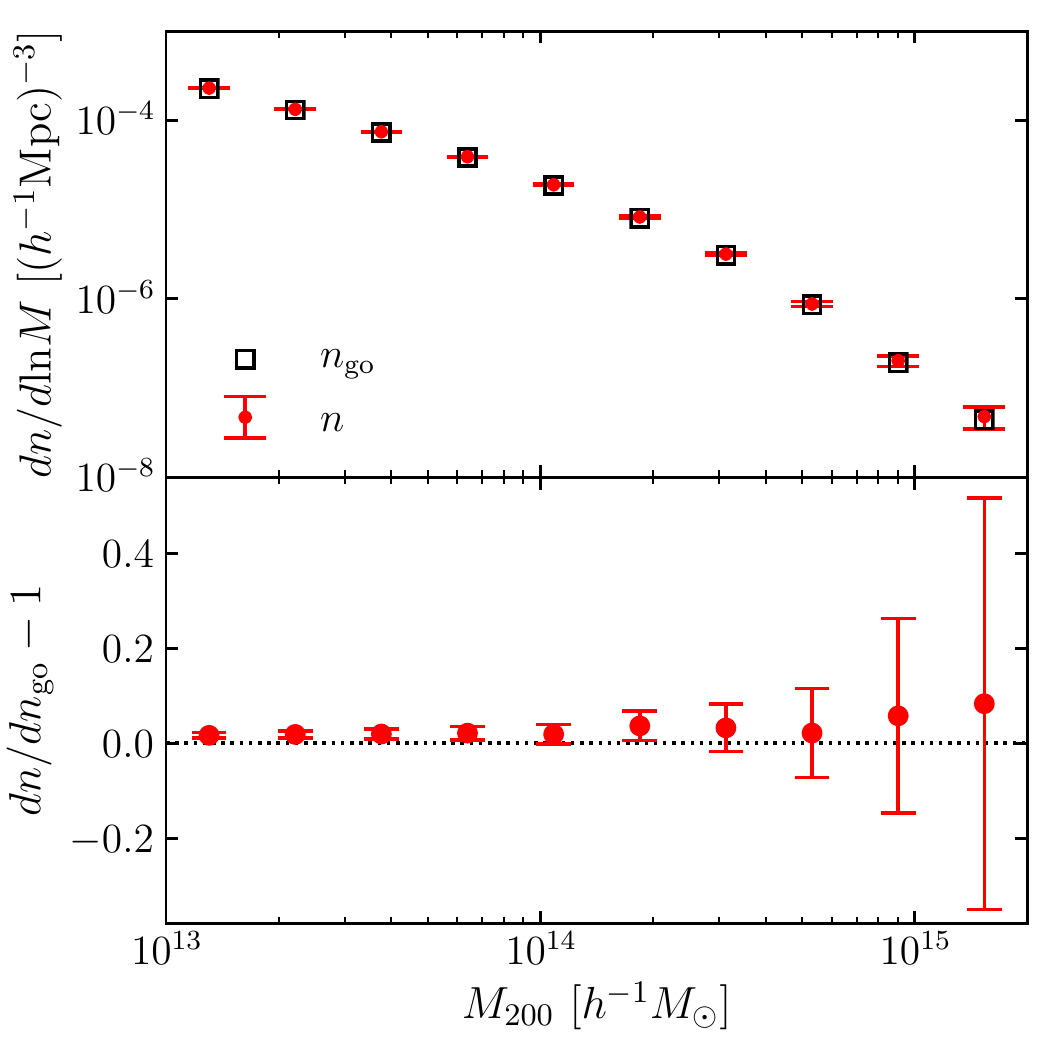}
\end{center}
\caption{Top panel compares the $z = 0$ mass function of SO halos from the 
GO (black squares) and NR (red circles) Borg Cube simulations. Error
bars denote the Poisson error in each mass bin and are shown only for the NR
simulation in the upper panel for the purpose of clarity. 
The bottom panel shows the relative difference between the two simulations.}
\label{fig:bcsodndm}
\end{figure}

The next useful statistic we study is the halo mass function.
Figure \ref{fig:bcsodndm} compares the SO mass function from the GO and NR
Borg Cube runs at $z = 0$. We observe the trend that the mass function in the NR
run is slightly enhanced for all masses within the range considered here.  
The change is rather modest with a nearly constant $2\%$ increase across the 
range $\unit{10^{13}}{\massh} \leq M_{200} \leq \unit{10^{14}}{\massh}$. This appears
to increase slightly for larger masses, though this enters the exponential tail of 
the mass function where the measurement is noisy. 

The interpretation is 
not that NR processes actually increase the number of massive halos, but
rather the mass definition is altered by the internal redistribution of matter
within individual halos. Since the two simulations use the same initial conditions,
we are able to identify counterpart halos between each run and test this directly.
Indeed, we find that halos in the NR run have $R_{200}$ that are $0.4\%$ 
larger on average than their GO counterparts; this equates to an average 
increase of $1\%$ in $M_{200}$. This number agrees well with the tSPH simulation 
of \citet{cui/etal:2012} who confirm that changes in the NR mass function 
can be accounted for using a simple shift in mass. This will also hold true once
cooling and feedback are considered though the change in halo mass will be much 
closer to the $10\%$ range \citep{stanek/etal:2009}.

\subsection{Baryon Fraction}

\begin{figure}
\begin{center}
\includegraphics[width=\smhwidth]{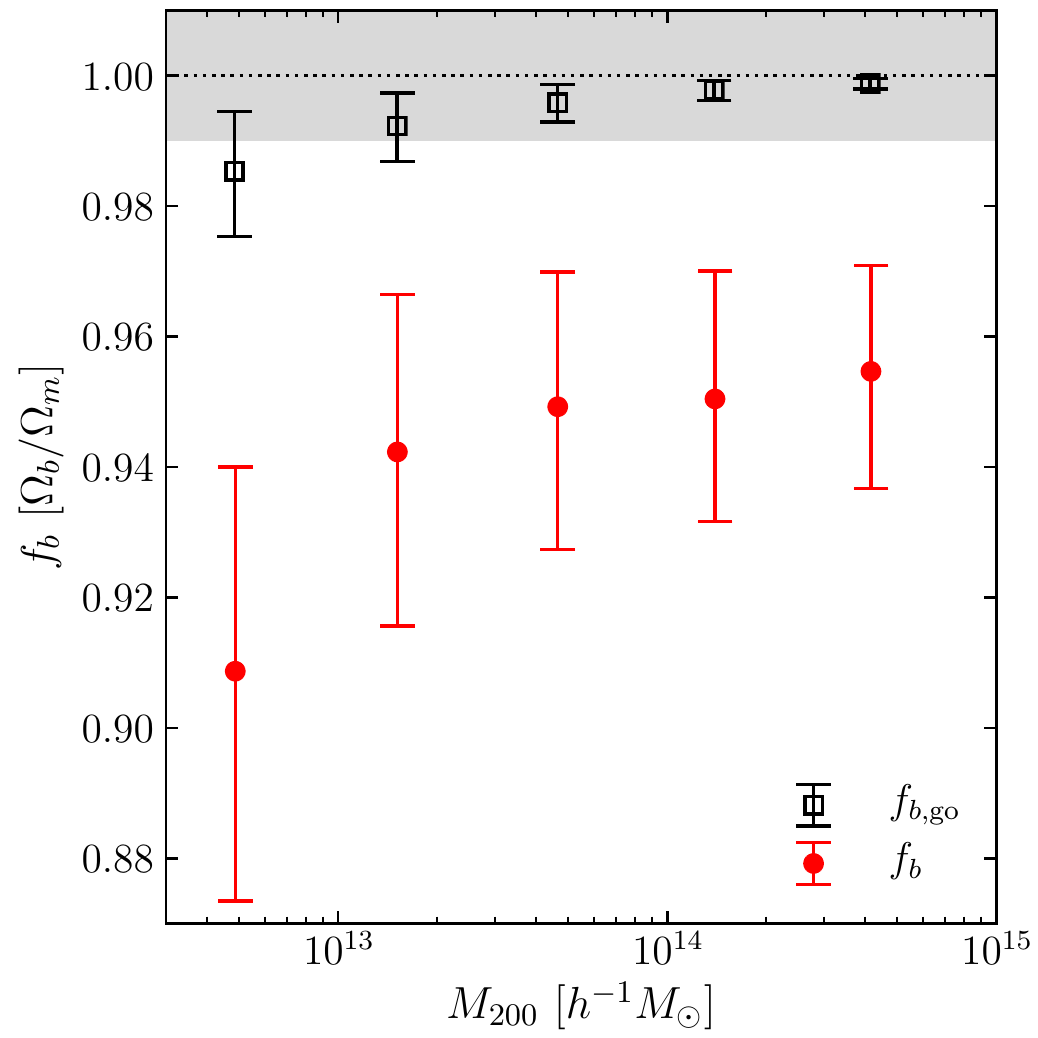}
\end{center}
\caption{Global baryon mass fractions in units of the universal mean computed for 
SO halos binned into five mass groups at $z = 0$. Black squares and red circles show 
the median value in each bin for the GO and NR Borg Cube simulations, respectively. 
In each case, error bars bracket the 25th and 75th percentiles of the distribution.
We naively expect the GO points to sit at unity and highlight $\pm1\%$ deviations
away from this using the shaded gray band. The departure of the GO result 
from unity indicates that the simulations are not converged for masses
$M_{200} \lesssim \unit{10^{13}}{\massh}$.}
\label{fig:fgasglobal}
\end{figure}

We now shift gears to focus on the relative distribution of baryons and CDM
within individual halos. The main quantity to explore is the global baryon mass fraction, 
$f_b \equiv M_{b,200}/M_{200}$, where $M_{b,200}$ is the baryon mass within a halo 
of total mass $M_{200}$. Figure \ref{fig:fgasglobal} shows $f_b$ in units of the 
universal baryon fraction for five bins in halo mass at $z = 0$. Red circles trace 
the median value in each bin for the NR run while error bars bracket the 25th
and 75th percentiles. We find that halos of mass $M_{200} \geq \unit{10^{13}}{\massh}$
have baryon fractions that are roughly about $95\%$ that of the cosmic mean. 
This lies slightly above the range of values ($f_b \approx 89\% - 94\%$) reported in 
previous NR simulations
\citep[e.g.,][]{ettori/etal:2006,crain/etal:2007,gottlober/etal:2007,rudd/etal:2008,
stanek/etal:2010,battaglia/etal:2013}
for similar masses. Intrinsic scatter in this quantity has been shown to exist amongst
hydro methods \citep{frenk/etal:1999,kravtsov/etal:2005} with the trend
that mesh-based codes tend to yield larger baryon fractions at the level of a few 
percent compared to tSPH. Hence, our results are more consistent with previous
mesh-based results.

The physical mechanism responsible for depleting the baryon fraction below the cosmic
mean is shock heating. As we will show later, this process tends to ``puff out'' 
baryons around $R_{200}$. This seems to be especially prominent for the lowest mass bin 
in Figure \ref{fig:fgasglobal} where $f_b$ drops to a level of $91\%$. Note that
\citet{gonzalez/etal:2013} find a weak dependence of decreasing baryon fraction with
decreasing $M_{500}$ in an observational study of a collection of galaxy clusters and
groups. Unfortunately, testing this dependence with low-mass halos 
in the Borg Cube becomes difficult since we expect resolution issues to 
emerge at some point as we push to smaller scales. This issue is most readily 
examined with the GO run.

As mentioned earlier, coupling between unequal-mass CDM and baryon particles
leads to artificially strong interactions on small scales. This will obviously
impact the central depths of halos and will be more prominent in low-mass 
systems where physical scales are smaller and potential wells shallower.
The black squares in Figure \ref{fig:fgasglobal} trace the global baryon fractions 
from the GO simulation. Ideally, we expect these numbers to match the universal value 
since there is no distinction between CDM and baryon particles in the absence of 
hydro forces. This appears to be the case for masses 
$M_{200} \geq \unit{10^{13}}{\massh}$ where $70\%$ of all halos are within $1\%$ 
of the cosmic mean. We see a clear trend of decreasing median and increasing
scatter as we move from high- to low-mass systems. For instance, $67\%$ of halos 
with mass $M_{200} < \unit{10^{13}}{\massh}$ have baryon fractions that deviate by more
than $1\%$ from the universal value. Based on this, we set 
$M_{\rm con} = \unit{10^{13}}{\massh}$ as the converged mass scale in the 
Borg Cube simulations. This is equivalent to roughly 3200 times the combined CDM 
and baryon particle mass. These findings agree well with the two-species
GO simulations of \citet{binney/knebe:2002,power/etal:2016} where halos containing fewer
than a few thousand particles are also shown to exhibit a strong deficit of the low-mass group.

\begin{figure}
\begin{center}
\includegraphics[width=\smhwidth]{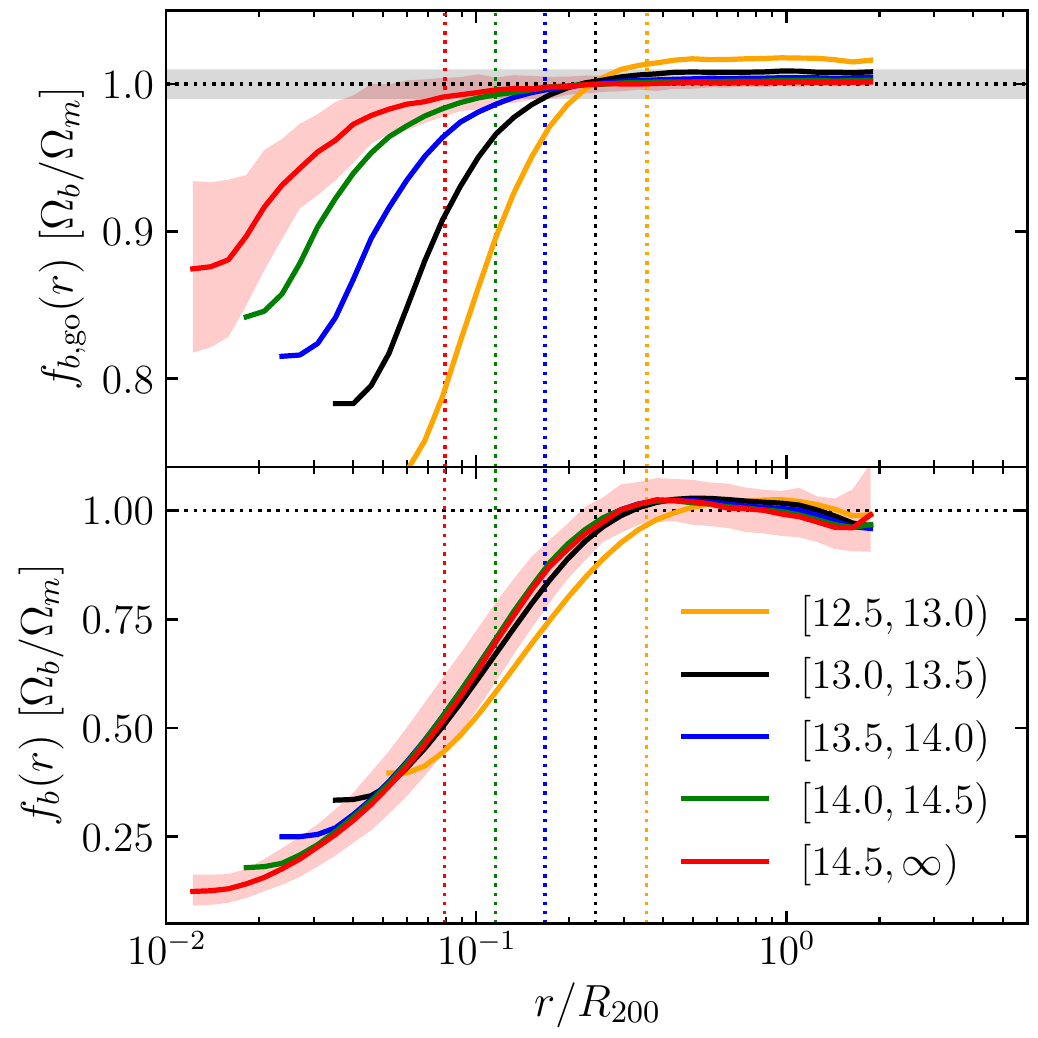}
\end{center}
\caption{The baryon fraction in radial shells normalized to the universal 
mean at $z = 0$ for the Borg Cube GO (top panel) and NR (bottom panel)
simulations. In each panel, the solid colored lines trace the stacked profiles
for halos in five mass groups as indicated by the legend showing the log$_{10}$ 
ranges of each bin. Each profile is truncated below the gravitational softening
length $r_{\rm soft} = \unit{14}{\kpch}$. The shaded gray band in the top panel
shows deviations of $\pm 1\%$ from unity, which we naively expect the GO result
to reside. The shaded red band in each panel brackets the 25th and 75th percentiles 
of individual profiles for the highest mass bin. Vertical dotted lines 
denote the location of the physical scale $r_{\rm con} = \unit{100}{\kpch}$ 
for the mass bin of the corresponding color.}
\label{fig:fgasradial}
\end{figure}

We can investigate the issue of convergence further by looking at radial profiles
of the baryon fraction. The top panel of Figure \ref{fig:fgasradial} 
shows this for the GO run where we overlay stacked profiles from the
same set of five mass bins. We first note that the four converged
mass bins are all within $1\%$ of the universal mean for $r \gtrsim 0.2R_{200}$.
The lowest-mass bin, in contrast, shows a noticeably higher baryon fraction
all the way out to $R_{200}$. In each case, as we move toward smaller radial bins,
we find a sharp drop in baryon fraction until eventually hitting a plateau as we
approach $r_{\rm soft}$. This sharp drop
seems to occur at a constant physical scale of $r_{\rm con} = \unit{100}{\kpch}$
as indicated by the vertical dotted lines showing where $r_{\rm con}$ intersects 
each mass bin. We attribute the precipitous drop in baryon fraction to artificially 
strong gravitational interactions within the central regions of halos leading to
strong mass segregation \citep{efstathiou/eastwood:1981}. Evidently, this process 
extends out to scales about seven times larger than the gravitational softening 
length. We conclude that 
individual baryon and CDM mass distributions in the GO run are converged down to
$r_{\rm con} = \unit{100}{\kpch}$. Recall that earlier we found the {\em total}
matter distribution to be converged down to smaller scales of about 
$r_{m,{\rm con}} = \unit{30}{\kpch}$ based on a comparison to the SS case 
(see Figure \ref{fig:powerrsoft}).

The bottom panel of Figure \ref{fig:fgasradial} shows analogous results from the
NR simulation. In this case, the profiles from each of the four converged 
mass bins sit roughly on top of each other, suggesting a universal form for halos 
$M_{200} \geq \unit{10^{13}}{\massh}$.
It is unclear whether such a universal form extends to lower masses. In contrast
to the GO run, we do not see any obvious features surfacing around the
scale $r_{\rm con}$. The reason is that this scale is overwhelmed by thermal 
pressure support that naturally reduces baryon clustering, thus alleviating 
the problems encountered in the GO run. Hence, convergence criteria in the
NR run are likely to be somewhat relaxed compared to the GO case. 
Note the same may not necessarily be true, for instance, in a hydrodynamic
simulation with cooling processes that tend to promote baryon clustering on small
scales, potentially exacerbating the issue further. 
While it is difficult to assess exactly to which scales the NR results are
converged here, the smoothness and universality of the curves in Figure 
\ref{fig:fgasradial} down to $r_{\rm soft}$ are encouraging.
We leave to future work a more
detailed investigation of the numerical interplay between multi-species 
gravitational interactions and hydrodynamic processes. 

\subsection{Concentration}

\begin{figure}
\begin{center}
\includegraphics[width=\smhwidth]{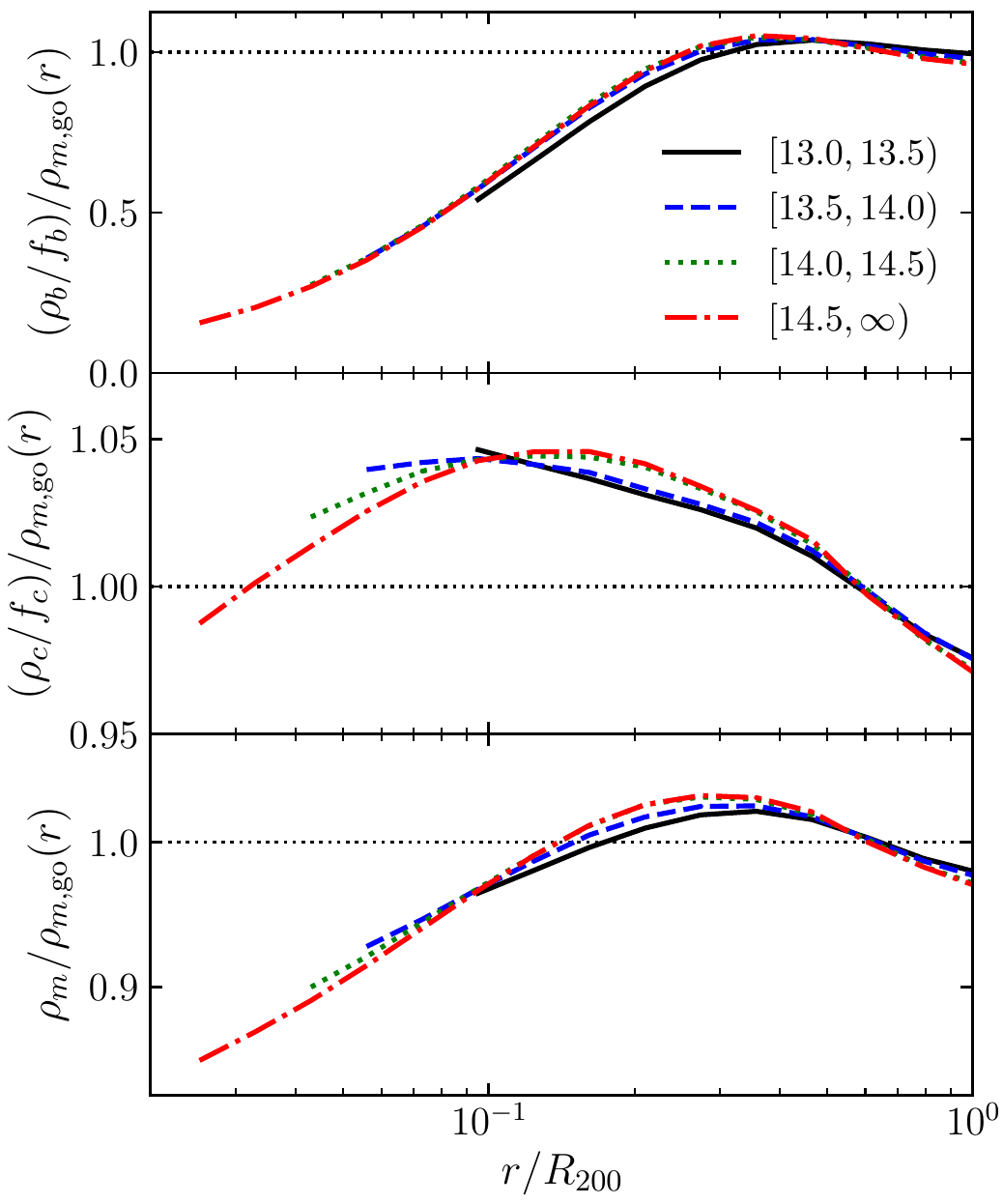}
\end{center}
\caption{Stacked radial density profiles of baryons (top), CDM (middle), and total
matter (bottom) in the NR run relative to the GO total matter profile
for each of the four converged mass bins at $z = 0$. Curves are truncated
below the total matter convergence scale of $r_{m,{\rm con}} = \unit{30}{\kpch}$
for the GO run.} 
\label{fig:rhorelative}
\end{figure}

Next we explore the internal redistribution of matter that arises in response
to shock heating. This complements the earlier analysis of the
power spectrum which can be derived by integrating over the distribution
of halo profiles via the halo model \citep[e.g.,][]{rudd/etal:2008}. 
In Figure \ref{fig:rhorelative} we compare 
radial density profiles from the GO and NR runs stacked over all halos in 
each of the four converged mass bins. The top, middle, and bottom panels show 
density profiles for baryons, CDM, and  total matter, respectively. 
In each case, we use the appropriately scaled total matter curve from the GO run as 
the comparison point and truncate each result below $r_{m,{\rm con}}$.

Figure \ref{fig:rhorelative} shows that changes in the individual species and 
total matter distributions are roughly consistent across all four mass bins. 
Unsurprisingly, baryons show the largest changes in the NR run with a slight
enhancement in density at $r \approx 0.4R_{200}$ followed by a sharp drop that
approaches one-tenth the density of the GO run at the halo center. Changes
to the CDM show qualitatively similar trends with a 
slight depression at $R_{200}$ followed by an increase that peaks around
$r \approx 0.2R_{200}$, and then a final small descent. Overall, the changes
in CDM are quite minor being within $\pm 5\%$ of the GO result across the
entire radial range. The total matter profile also starts with a small decrement 
at $R_{200}$ followed by an upturn that peaks at $r \approx 0.3R_{200}$ and finishes
with a suppression that approaches $85\%$ of the GO density at the halo center.
This result is consistent with the cumulative mass profiles from the modern SPH
and mesh-based codes used in the nIFTy comparison project \citep{cui/etal:2016}.

\begin{figure}
\begin{center}
\includegraphics[width=\smhwidth]{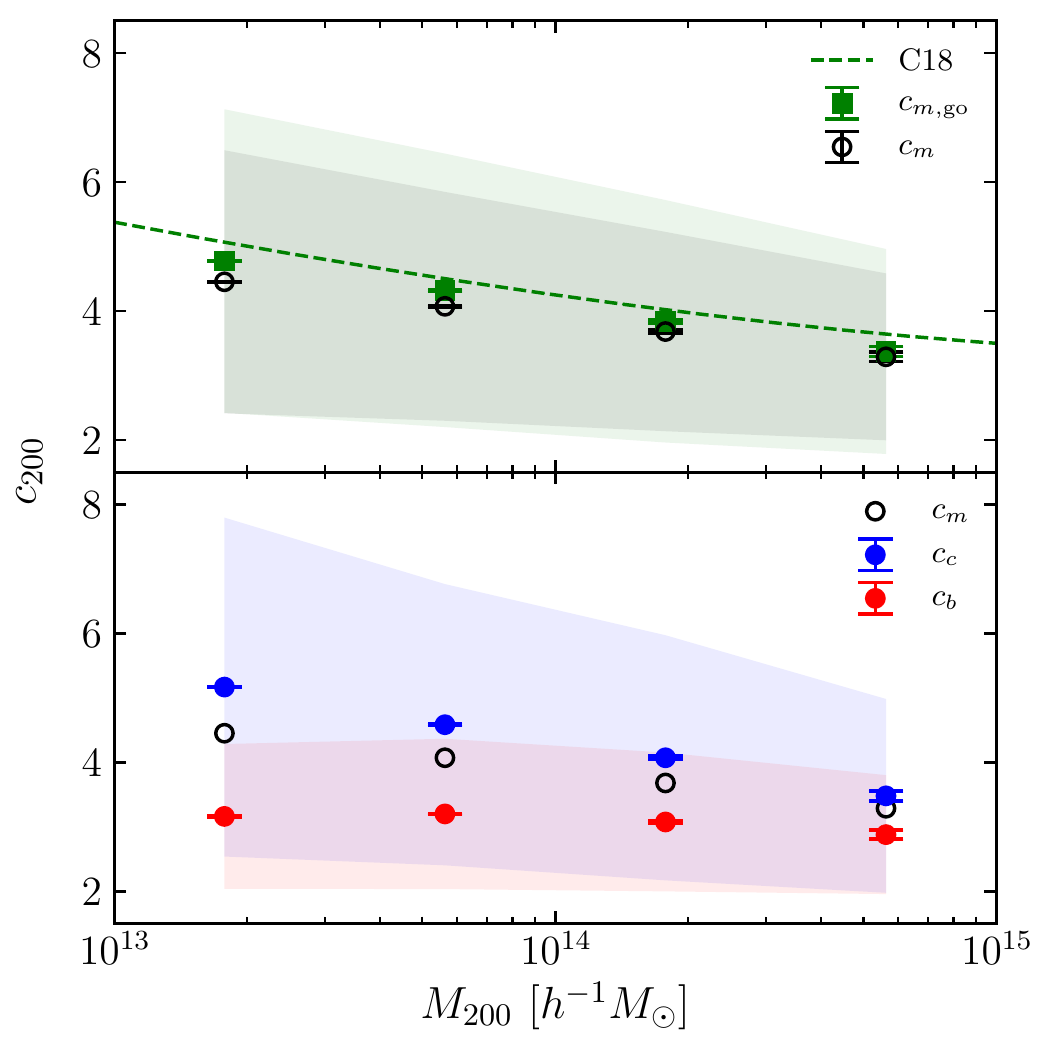}
\end{center}
\caption{Concentration-mass relation for all SO halos at $z = 0$. Top panel
compares $c_{200}$ based on the total matter profiles from the GO (green
squares) and NR (black circles) Borg Cube runs. In each case, the data point
shows the mean value of $c_{200}$ in each mass bin with the error bars denoting
the standard error of the mean and the shaded region highlighting the standard
deviation of the bin. The dashed green line traces
the corresponding \citet{child/etal:2018} fitting function. The bottom panel
focuses on the individual baryon and CDM components from the NR run. Blue
(red) circles follow the mean $c_{200}$ for CDM (baryons) with the black
circles repeating the NR total matter relation from the top panel for 
comparison.}
\label{fig:concentration}
\end{figure}

One way to summarize the redistribution of matter within halos is through the
concentration. We plot the concentration-mass relation for all SO halos at $z = 0$
in Figure \ref{fig:concentration} using the ``peak finding'' concentration
method described earlier. The top panel shows $c_{200}$
computed from the total matter profiles for the GO and NR runs. For comparison,
the dashed green line traces the \citet{child/etal:2018} fitting function based
on individual profiles for all (relaxed and unrelaxed) halos. This is in good
agreement with the GO run which uses the same cosmology for which the fitting 
function was calibrated.

Comparing the total matter concentrations from the
two runs shows a clear trend in a reduction of $c_{200}$ in the NR case. 
The difference is stronger at smaller masses with a $7\%$ reduction in the lowest
mass bin and only a $2\%$ reduction in the highest mass bin. This disagrees with some
earlier works \citep{rudd/etal:2008,rasia/etal:2013} that find a $5-10\%$ 
{\em increase} in NR concentration over the same mass range. This discrepancy
is not related to the definition of concentration since we find a similar
level of reduction in $c_{200}$ when using concentrations based on fits
to NFW profiles. Rather, the difference is likely sourced by a combination 
of two effects: 1) the earlier works were based on much smaller samples of halos 
with limited statistics; 2) the measurement of concentration is affected by
the redistribution of baryons on small scales, which depends on the
specific hydro solver, as evidenced in the measurements of power spectra
in Figure \ref{fig:bcpower}.

The bottom panel of Figure \ref{fig:concentration} compares the concentrations
from the baryon and CDM components of each SO halo in the NR run. 
As expected, baryon values are lower than CDM due to their shallow density 
profiles. Interestingly, the baryon result is nearly independent of mass with 
mean values $c_{200} \approx 3$ for each mass bin. CDM concentrations are 
markedly larger than baryons and display the usual trend of increasing
$c_{200}$ with decreasing mass. Moreover, we find the CDM concentration 
in the NR run is larger than the GO total matter concentration at a level of 
$3\%$ for the highest mass bin and $8\%$ for the lowest mass bin. This reflects
the earlier observations of increased small-scale CDM power 
(Figure \ref{fig:bcpower}) and enhanced density near the scale radius (middle
panel of Figure \ref{fig:rhorelative}).

While comparisons with earlier simulations are somewhat limited, this analysis
shows that the predicted change in total matter concentration induced by
NR processes depends on the details of the hydro solver. Of course, the inclusion of
additional physics will significantly alter the density profiles
and concentration-mass relations seen here. Cooling, on
the one hand, increases concentration as baryons condense in the core while
feedback, on the other hand, decreases concentration as matter is expelled from
the inner to outer regions of a halo 
\citep{rasia/etal:2013,shirasaki/etal:2018}.


\section{Gas Profiles}
\label{sec:gprof}

\begin{figure}
\begin{center}
\includegraphics[width=\smhwidth]{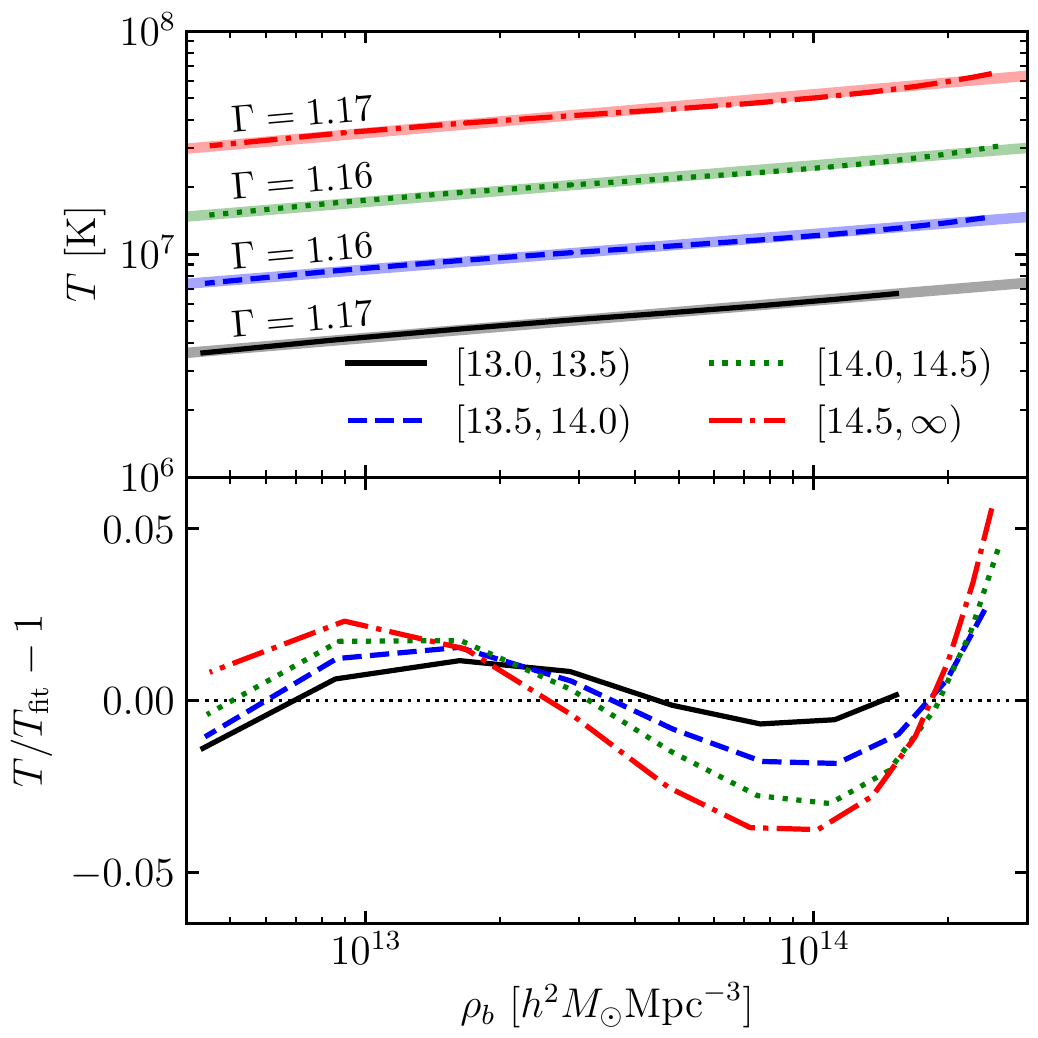}
\end{center}
\caption{Stacked temperature versus density profiles for each of the four SO halo
mass bins in the NR run at $z = 0$. Each of the lightly shaded solid lines in
the top panel traces the linear regression performed on the mass bin of the 
corresponding color. Adding one to the slope of each line yields the polytropic index
which ranges from $1.16-1.17$ for the different mass bins. The bottom panel
shows the relative difference between the stacked profiles and the linear
regression.}
\label{fig:gamma}
\end{figure}

We have seen that NR processes significantly alter the internal structure
of halos. We proceed here with a more complete description of the gaseous
component of halos by measuring density, temperature, entropy, and pressure
profiles over a wide range in mass. This analysis can be used in
conjunction with analytic treatments like the halo model to construct a synthetic 
picture of the cosmological gas distribution. For this purpose, we also compare our
results to a simple model based on the idea that gas exists in hydrostatic
equilibrium (HSE) within the potential well of its host.
Of course, such a description is only an idealized approximation since many halos
are unrelaxed and therefore not in a state of equilibrium. In any event, it is useful to test the
predictive power of HSE.
In what follows, we work with the \citet{komatsu/etal:2001} model based on 
the assumptions that baryons 1) trace the total matter density profile on halo
outskirts, 2) follow a polytropic equation of state, and 3) exhibit full thermal
pressure support. The first assumption is justified by the baryon fractions 
hovering around unity at $R_{200}$ in Figure \ref{fig:fgasradial} 
while the second assumption will be justified in Figure \ref{fig:gamma}. 
We will explore the validity of the third assumption in the next section.

We begin with an overview of the HSE model compared to here. We point the interested 
reader to \citet{komatsu/etal:2001} for a more careful derivation and also to
\citet{rabold/teyssier:2017} who perform a similar analysis to our own.
The starting point is based on the idea that the total matter distribution 
follows an NFW form:
\bq
\rho_m(x) = \frac{\rho_s}{cx(1+cx)^2},
\label{eq:nfw}
\eq
where $x \equiv r/R_{200}$, $c$ is the concentration, and $\rho_s$
is a characteristic density set by the concentration:
\bq
\rho_s = \frac{c^3\rho_{200}}{3f(c)}.
\label{eq:rhos}
\eq
Here $\rho_{200}$ is the overdensity criteria of the SO halo (i.e., 200 times 
the critical density) and 
\bq
f(x) = {\rm ln}(1+x) - \frac{x}{1+x}.
\label{eq:fx}
\eq
If we assume the baryons follow a polytropic equation of state,
$P \propto \rho_b T \propto \rho_b^\Gamma$, with pressure $P$, temperature $T$,
and polytropic index $\Gamma$, we can derive the density profile that results from
HSE within an NFW potential:
\bq
\rho_b(x) = \rho_0 
\left[ \frac{{\rm ln}(1+cx)}{cx} \right]^{{1}/({\Gamma-1})}.
\label{eq:heq-rhob}
\eq
Here $\rho_0$ is the asymptotic density approached at the halo center, and
is constrained by requiring that the baryon density trace the total matter
density at $R_{200}$. Setting $\rho_b(1) = f_b\rho_m(1)$ yields:
\bq
\rho_0 = \frac{f_b\rho_s}{c(1+c)^2}
\left[ \frac{{\rm ln}(1+c)}{c} \right]^{{-1}/({\Gamma-1})}.
\label{eq:heq-rho0}
\eq
The temperature profile follows as:
\bq
T(x) = T_0 \frac{{\rm ln}(1+cx)}{cx},
\label{eq:heq-temp}
\eq 
where the central temperature is derived from the requirement that the 
density profile drop to zero at infinity:
\bq
T_0 = \frac{4\pi G \mu m_p \rho_s R_{200}^2}{k_B c^2}\frac{\Gamma-1}{\Gamma}.
\label{eq:heq-t0}
\eq 
Here $G$ is the gravitational constant, $k_B$ the Boltzmann constant, 
$\mu = 0.59$ the mean molecular weight for a fully ionized gas, and 
$m_p$ the proton mass. Finally, the pressure follows as:
\bq
P(x) = \frac{\rho_0}{\mu m_p}k_B T_0
\left[ \frac{{\rm ln}(1+cx)}{cx} \right]^{{\Gamma}/({\Gamma-1})}.
\label{eq:heq-pressure}
\eq 

The preceding derivation has two free parameters: 1) the polytropic index
and 2) the concentration of the total matter density profile. We compute the
polytropic index by performing a linear regression on a log-log plot of
temperature versus density. This is shown in Figure \ref{fig:gamma} for each
of the four mass bins. The best-fit values of $\Gamma$ range from 1.16 to 1.17, 
which is similar to those found in earlier works
\citep{komatsu/etal:2001,ascasibar/etal:2003,rabold/teyssier:2017}. Though the
assumption that a single value of $\Gamma$ holds across the entire radial range
is only an approximation \citep{kay/etal:2004,battaglia/etal:2012b}, 
the deviations seen here from the best-fit constant values are relatively small.
The most noticeable difference shows up as an upturn in the $T-\rho_b$ relation 
at high density suggesting that larger values of $\Gamma$ may be more appropriate
in the central regions where $r/R_{200} \lesssim 0.1$. For the sake of
simplicity, we will use a constant value of $\Gamma = 1.17$ for each of the 
mass bins in the following analysis.

\begin{figure}
\begin{center}
\includegraphics[width=\smhwidth]{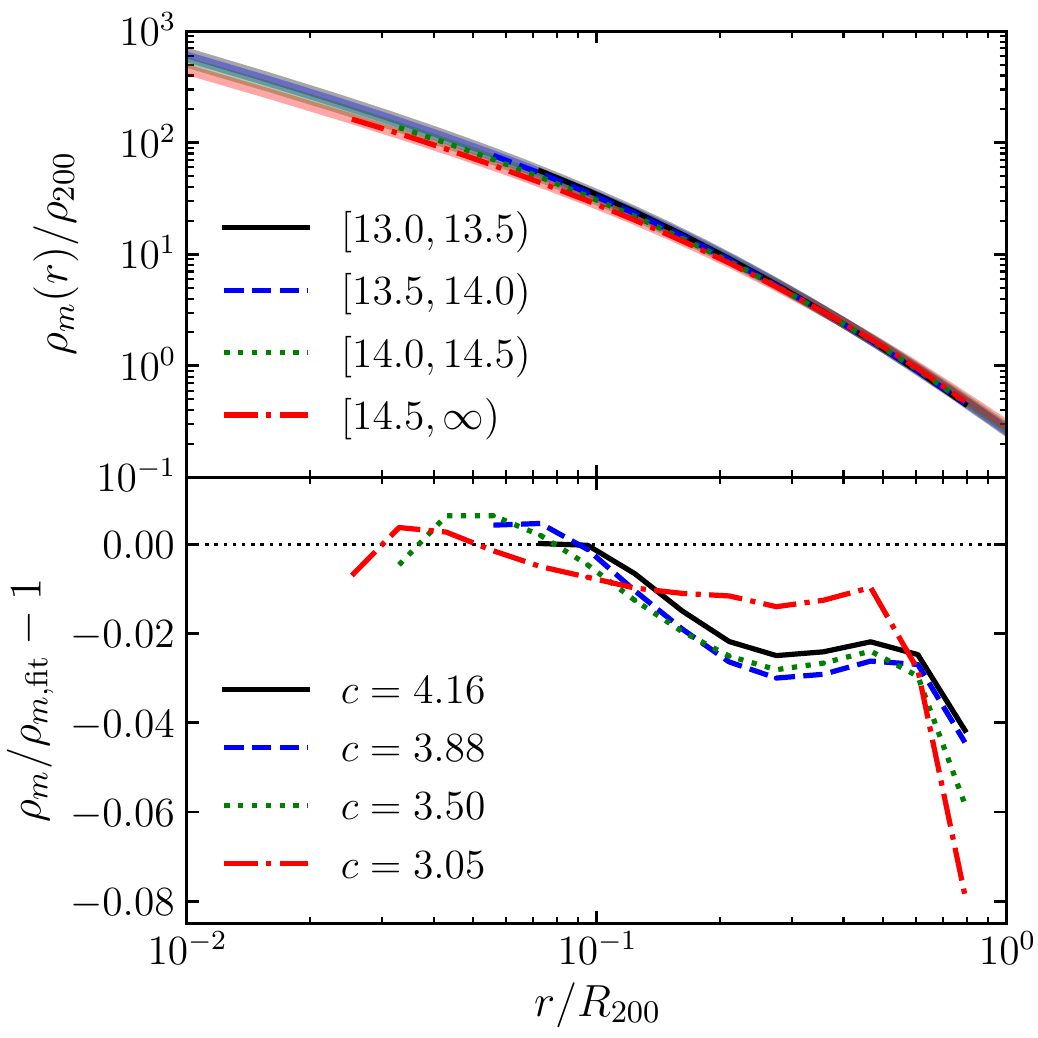}
\end{center}
\caption{Stacked total matter density profiles for each of the four SO halo
mass bins in the NR run at $z = 0$. Each stacked profile is truncated below
twice the gravitational softening length and above $R_{200}$. The lightly
shaded solid lines in the top panel trace the best-fit NFW profiles for
the mass bin of the corresponding color. The bottom panel shows the relative 
difference between the stacked profiles and the NFW fit with the concentration 
of each mass bin listed in the legend.}
\label{fig:fitrhom}
\end{figure}

We determine the NFW concentration of the total matter distribution by fitting
equation (\ref{eq:nfw}) to the stacked density profile of each mass bin. The
results are shown in Figure~\ref{fig:fitrhom}. When performing the fit, we
avoid resolution issues by truncating the stacked density profile below twice
the softening length, and also include only those radial bins up to $R_{200}$.
Overall, an NFW profile does a relatively good job at describing the stacked 
total matter profiles with deviations within $3\%$ over most of
the radial range. The deviations seem to get worse with increasing
radial distance with the stacked profiles systematically falling off more steeply 
than an NFW form. This trend was also observed in \citet{child/etal:2018},
and is not surprising given that we are stacking over relatively wide mass bins.
In any event, the deviations found here are relatively small, and we compute the 
best-fit NFW concentrations $c = 3.05$, 3.50, 3.88, and 4.16 for the four bins in 
descending order of mass. 

\begin{figure*}
\centering
\begin{tabular}[b]{@{}p{0.48\textwidth}@{}}
\centering\includegraphics[width=\smhwidth]{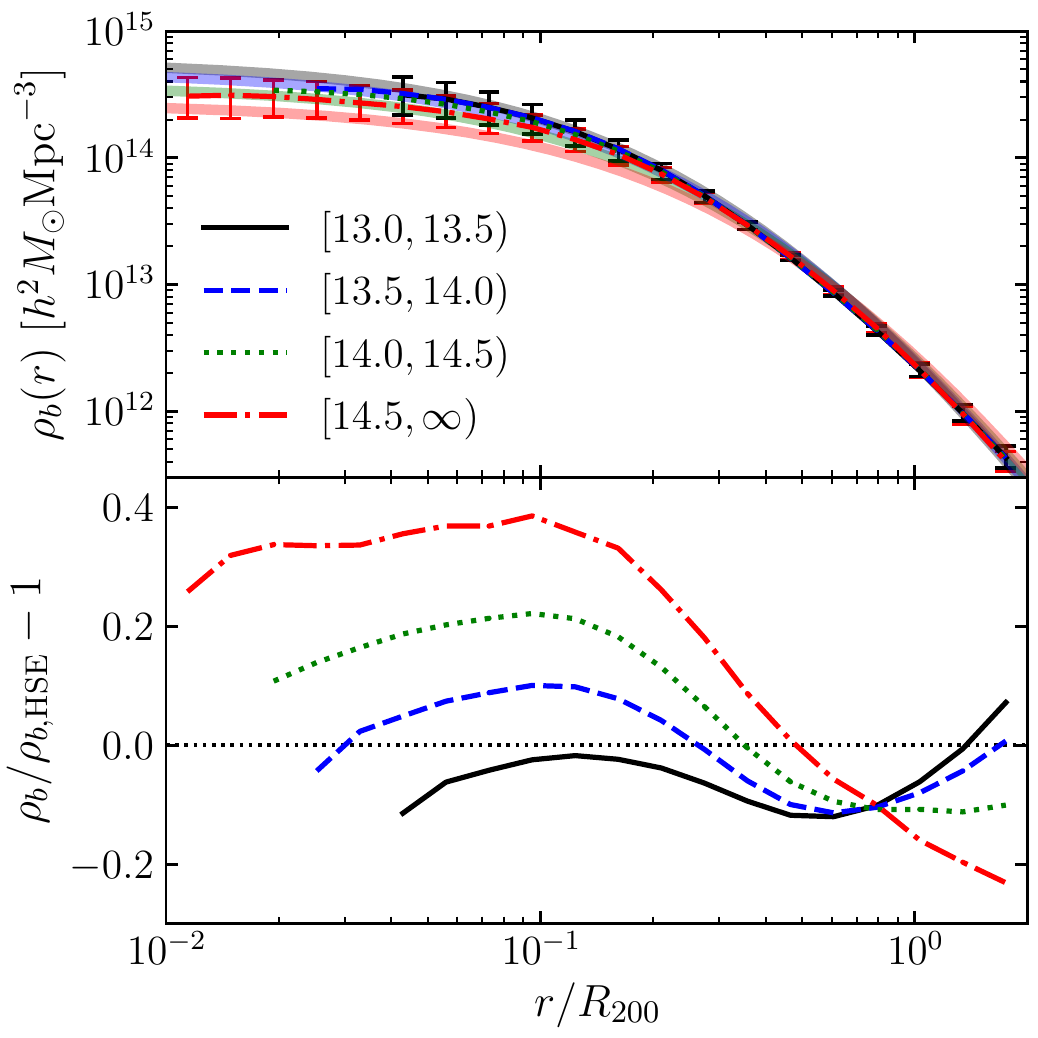}\\[0.5cm]
\centering\includegraphics[width=\smhwidth]{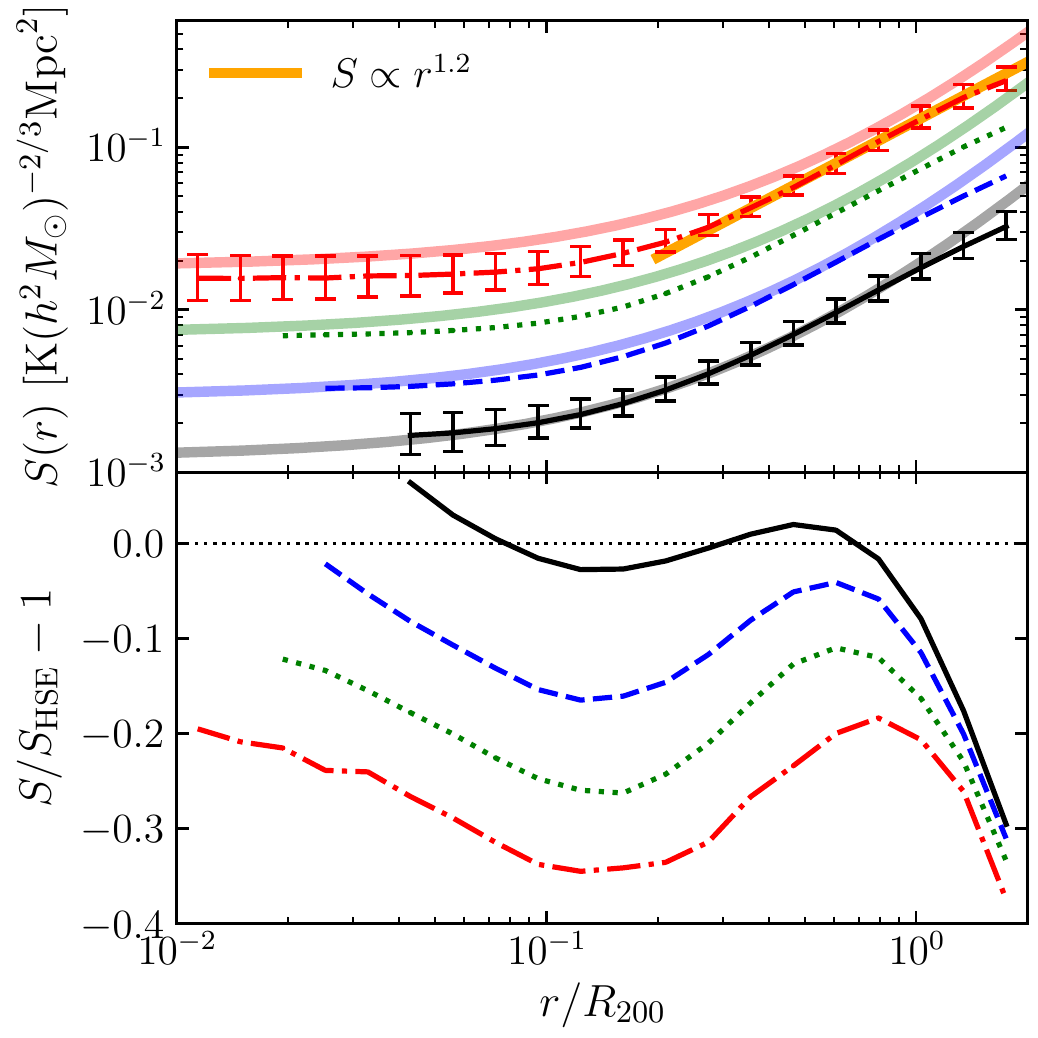}
\end{tabular}%
\quad
\begin{tabular}[b]{@{}p{0.48\textwidth}@{}}
\centering\includegraphics[width=\smhwidth]{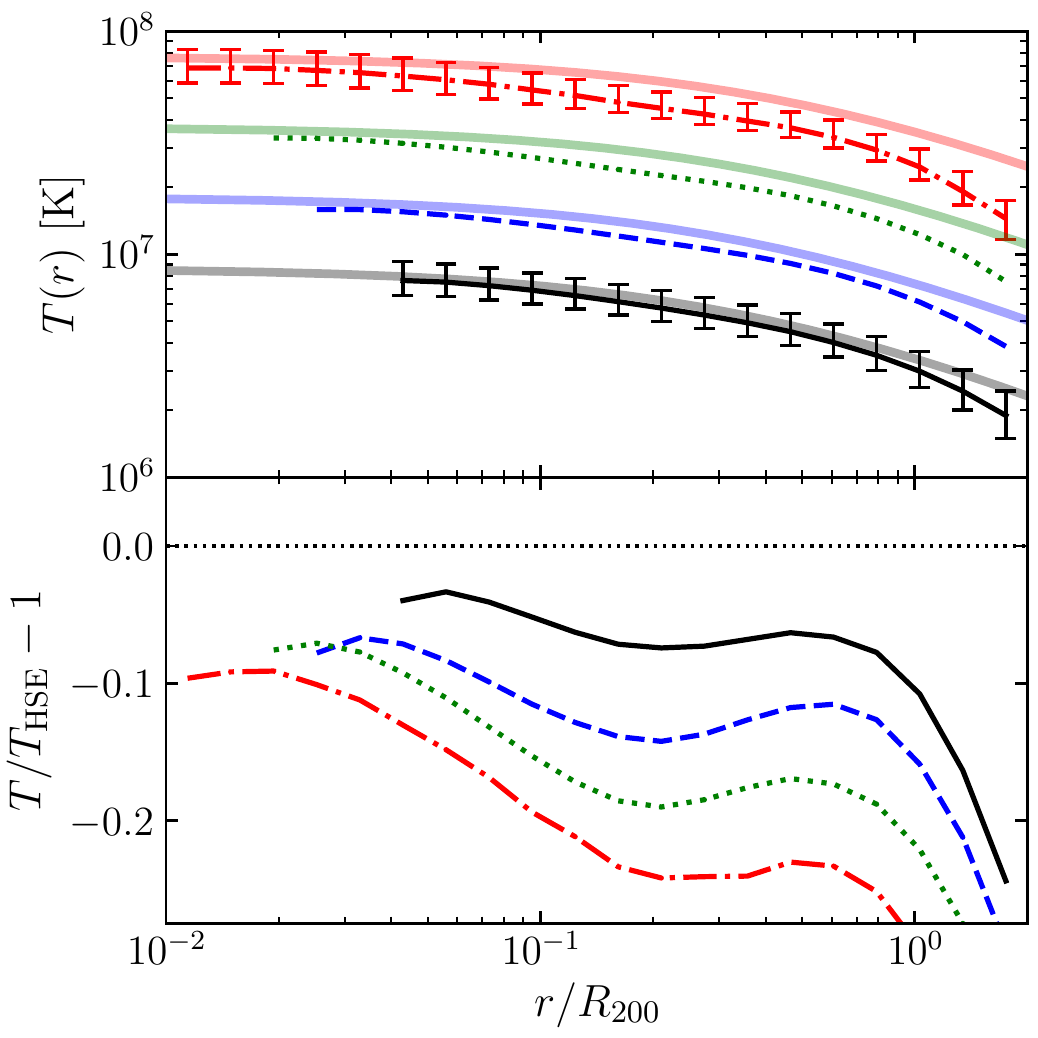}\\[0.5cm]
\centering\includegraphics[width=\smhwidth]{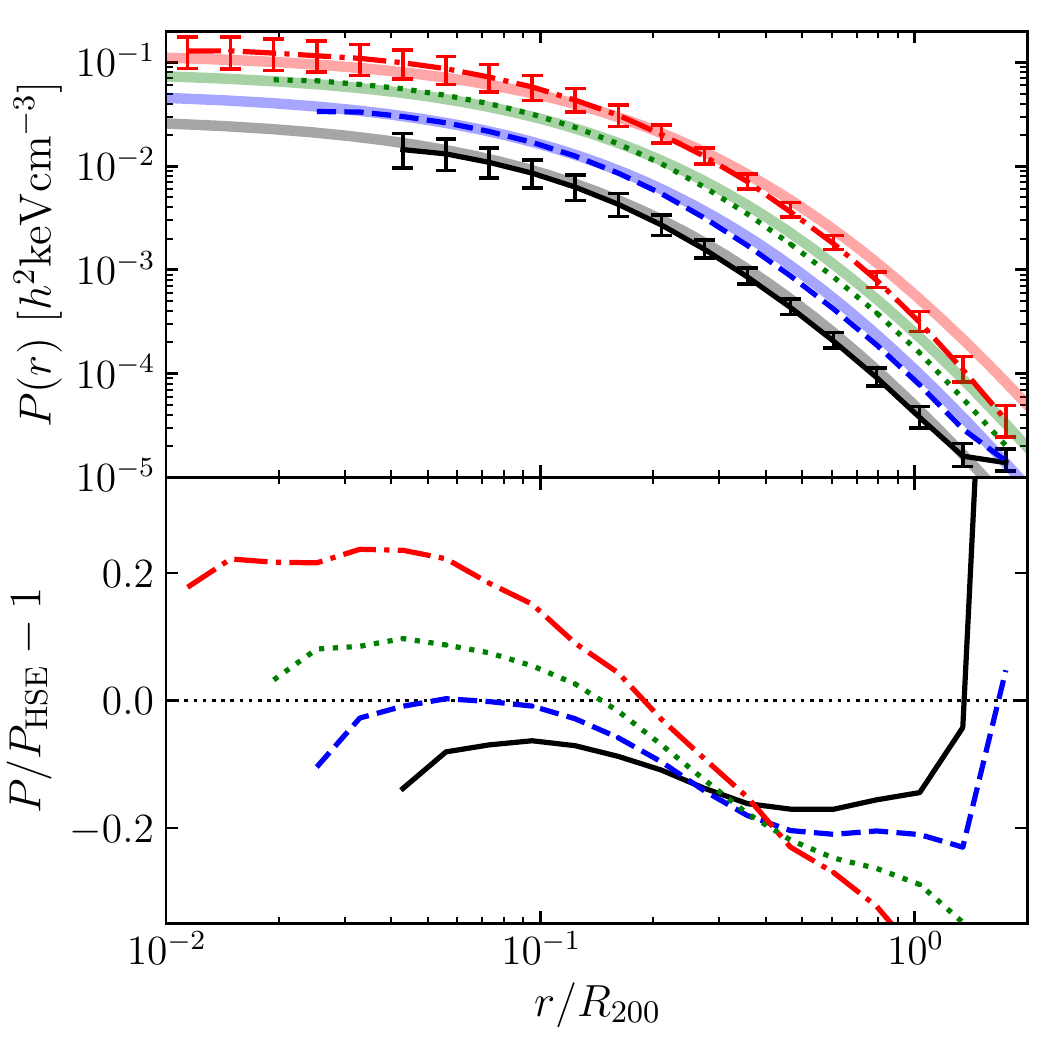}
\end{tabular}
\caption{Stacked profiles for density, temperature, pressure, and entropy in
clockwise order from the top-left. Opaque lines in the upper panel of each plot show
the NR result at $z = 0$ with the profile for each mass bin truncated below the
gravitational softening length. Stacks use the median value in each bin and we show
the 25th and 75th percentiles of the highest (lowest) mass bin with red (black)
error bars. The error bars do not depend strongly on mass and are thus only shown for
the highest and lowest bins to avoid overcrowding the plots.
The lightly shaded lines trace the HSE prediction for the mass bin of the corresponding
color. The lower panel of each plot shows the relative difference between the simulation 
curves and the HSE model. The thick orange line in the entropy plot shows the
$S \propto r^{1.2}$ power-law fit to the highest mass bin in the range $r > 0.2R_{200}$.}
\label{fig:hyeqbm}
\end{figure*}

Figure~\ref{fig:hyeqbm} shows stacked profiles for density, temperature, 
entropy, and thermal pressure for each of the four mass bins. The opaque lines in the 
upper panel of each plot trace the median value of the stacked profile while the red (black)
error bars show the 25th and 75th percentiles of the highest (lowest) mass bin.
The error bars are relatively insensitive to mass so we show them for only two mass bins
to avoid overcrowding the plots. The lightly 
shaded lines show the HSE prediction of the mass bin of the corresponding color using a \
constant $\Gamma = 1.17$ and the best-fit NFW concentration from the total matter 
density profile. The lower panel in each plot shows the relative difference between 
the stacked profile and the HSE prediction. 

We begin by looking at density in the top-left panel of Figure \ref{fig:hyeqbm}.
In contrast to the cuspy NFW profiles typical of CDM, baryons asymptote to
relatively flat central density cores for $r/R_{200} \lesssim 0.1$. This trend
is consistent with that observed in mesh-based and modern SPH simulations
\citep[e.g.,][]{frenk/etal:1999,sembolini/etal:2016}, and also arises naturally in the
HSE model. The three lower-mass bins agree within $20\%$ of the HSE model for 
the entire radial range, while the highest-mass bin has maximum deviations occurring at 
the $40\%$ level. In general, it appears as though the HSE model becomes more inaccurate
with increasing mass. This could reflect the fact that higher mass halos
are more likely to be unrelaxed so that the assumption of HSE is less valid.
The HSE curves all systematically underestimate the density 
at $R_{200}$, which follows from the trend in Figure \ref{fig:fitrhom} that the 
stacked profiles fall off steeper than NFW near the halo radius. The intersection
of the four curves in the lower panel arises from the fact that both the simulated
and model curves pass through each other around $r/R_{200} \approx 0.7$. Such a
feature is expected from equation (\ref{eq:heq-rhob}) which exhibits a pivot at
$x = 0.7$ with changing concentration.

The top-right panel in Figure~\ref{fig:hyeqbm} shows stacked temperature profiles.
The separate mass bins show self-similar results with inwardly rising temperatures
that approach a core value roughly three times greater than the value at $R_{200}$.
Indeed, \citet{loken/etal:2002} confirmed that a universal temperature profile
arises within NR simulations. The HSE model does a good job at predicting the
core temperature from the simulation, but heavily overestimates its value at $R_{200}$. 
A similar trend was noticed in \citet{rabold/teyssier:2017} who suggest a modification 
to equation (\ref{eq:heq-temp}) that accounts for contributions from turbulent pressure
support. Their correction increases with radius and reduces the HSE prediction by about
$6\%$ at $R_{200}$. Even without this correction, the HSE model does a relatively
decent job at matching the simulation, being within about $30\%$ for $r < R_{200}$ in
each mass bin. We will explore later the issue of non-thermal pressure support.

Next we examine entropy, which we define as $S \equiv T \rho_b^{-2/3}$.
Entropy is a particularly useful quantity since it plays the fundamental role in 
shaping the density and temperature distributions. It also encapsulates the 
thermodynamic history of heating and cooling processes\footnote{We remind
the reader that no cooling processes occur in our NR simulation here.}
during structure formation
\citep[e.g.,][]{voit:2005}.
The simulation curves in the bottom-left panel of Figure
\ref{fig:hyeqbm} show self-similar behavior with inwardly decreasing entropy that 
approaches an isentropic core with a value about one-tenth that at $R_{200}$. 
\citet{voit/etal:2005} showed that entropy profiles of NR halos follow a 
power-law with slope 1.2 for $r > 0.2R_{200}$. We also recover this trend, as seen
with the thick orange line comparing this power-law to the highest mass bin.
Coincident with our previous results, the HSE model matches the low-mass bins quite 
well and stays within $40\%$ for all bins up to $R_{200}$. The systematic 
over-prediction of entropy for $r \gtrsim R_{200}$ follows from the 
breakdown of the assumption of full thermal pressure support. As shown
in the next section, contributions from non-thermal pressure support increase
with radius, possibly explaining the sharp drop-off in agreement between the
simulation entropy and the HSE prediction seen in the highest radial bins.

Finally, the bottom-right panel of Figure \ref{fig:hyeqbm} shows stacked
thermal pressure profiles. As before, we see self-similar results across
each mass bin with inwardly rising pressure that increases by almost three
orders of magnitude in the central region compared to $R_{200}$. Self-similar
cluster pressure profiles have previously been found in both simulations and
observations \citep{nagai/etal:2007b,arnaud/etal:2010}. The simulation
results agree within about $20\%$ of the HSE model for $r/R_{200} \leq 0.4$,
but fall systematically below on larger scales. Again, this is a consequence
of the fact that a significant fraction of pressure support on those scales
is non-thermal. The flattening of the simulated 
pressure profiles for the two lower mass bins above $R_{200}$ is likely the
result of contributions from the surrounding environment.

In summary, we find a great deal of self-similarity in the thermodynamic properties
of the gas over a wide range in halo mass. This arises because the range in concentration
seen here is rather small and NR processes are scale free. This trend will likely
break down with the inclusion of more sophisticated gas treatments including cooling, 
star formation, and feedback which vary with halo mass. These processes most strongly
impact the central regions of halos so the results presented here will be
most applicable to halo outskirts. In fact, \citet{burns/etal:2010} found
that NR simulations are sufficient at matching the density, temperature, and
entropy profiles of observed clusters on scales $r \gtrsim 0.5R_{200}$. Similar
conclusions have been drawn by comparing NR simulations to those with
additional physics
\citep[e.g.,][]{loken/etal:2002,roncarelli/etal:2006,eckert/etal:2012,planelles/etal:2014,
cui/etal:2016,shirasaki/etal:2018}. 


\section{Hydrostatic Masses and Non-thermal Pressure}
\label{sec:hse}

The preceding analysis showed that the simple analytic model of
\citet{komatsu/etal:2001} was able to match the gas distribution of Borg Cube
halos to within $40\%$ for halos spanning two orders of magnitude in mass.
The largest discrepancies in the HSE model occurred close to $R_{200}$ and
are at least partly explained by its omission of non-thermal pressure support, which
we show later to be most important at large 
radial distance and high halo mass.
The major contribution of non-thermal support is expected in the form of 
kinetic pressure from turbulent gas motions and bulk flows that generally 
increase with halo-centric radius.
This accounts for $\approx 10-20\%$ of the total pressure support and leads 
to a bias of the same magnitude when deriving cluster masses based on HSE
\citep{evrard/1990,kay/etal:2004,faltenbacher/etal:2005,rasia/etal:2006,
nagai/etal:2007a,jeltema/etal:2008,piffaretti/valdarnini:2008,lau/etal:2009,
battaglia/etal:2012a,nelson/etal:2012}.
Obviously, this is an important effect to understand and the large volume
of the Borg Cube allows us to study its impact over a wide range in halo mass.

\begin{figure}
\begin{center}
\includegraphics[width=\smhwidth]{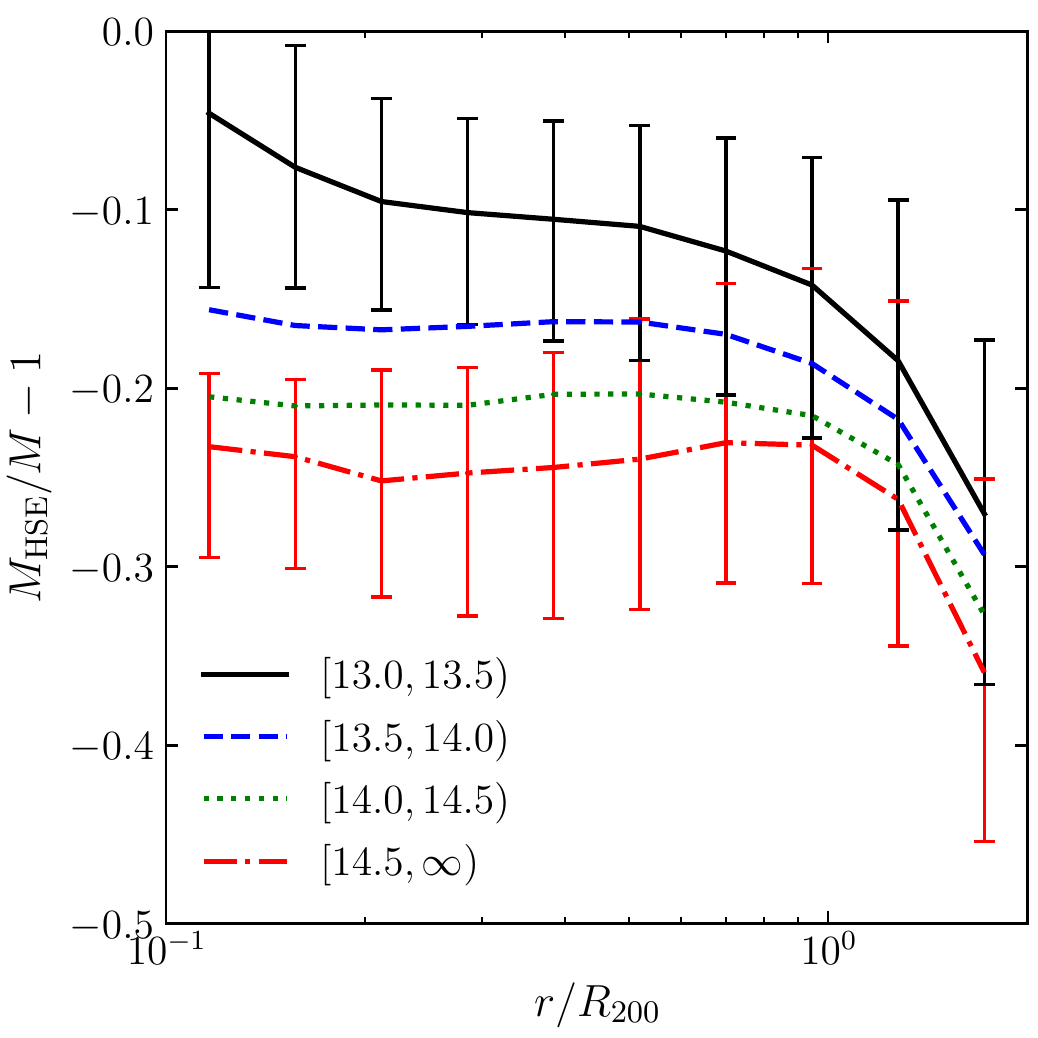}
\end{center}
\caption{Bias between the HSE mass estimate and simulation mass at $z = 0$.
Each line traces the median relation with the red (black) error bars showing
the 25th and 75th percentiles of the highest (lowest) mass bin. We show the result down to only
$0.1R_{200}$ as the relation becomes dominated by noise on smaller scales.}
\label{fig:mbias}
\end{figure}

The assumption of HSE with spherical symmetry and full thermal pressure support
can be used to estimate the mass of a galaxy cluster. The mass contained within
radius $r$ is computed as:
\bq
M_{\rm HSE}(<r) = -\frac{k_BT(r)r}{G\mu m_p}
\left[ \frac{{\rm d\ ln}\rho_b(r)}{{\rm d\ ln}r} + 
\frac{{\rm d\ ln}T(r)}{{\rm d\ ln}r} \right].
\label{eq:heq-mass}
\eq 
We use this expression to compute a radial HSE mass estimate for each halo and 
compare this to the mass profile measured directly from the simulation.
The resulting mass bias is shown in Figure \ref{fig:mbias}. For the most massive
bin, the HSE mass is biased low at a constant level of about $24\%$. This bias
drops with decreasing halo mass with the
mean bias in the radial range $[0.1-1]R_{200}$ being $21\%$, $17\%$, and
$10\%$ for the three lower mass bins. These findings are in good agreement with 
the $\approx20\%$ HSE mass biases for $M_{200} \gtrsim \unit{10^{14}}{\massh}$ halos
found in previous NR simulations \citep{kay/etal:2004,nelson/etal:2012}.
Here we have assumed full information on both the density and temperature 
profiles in equation (\ref{eq:heq-mass}). Incomplete knowledge on either
of these quantities has the potential to further bias the HSE estimate.

\begin{figure}
\begin{center}
\includegraphics[width=\smhwidth]{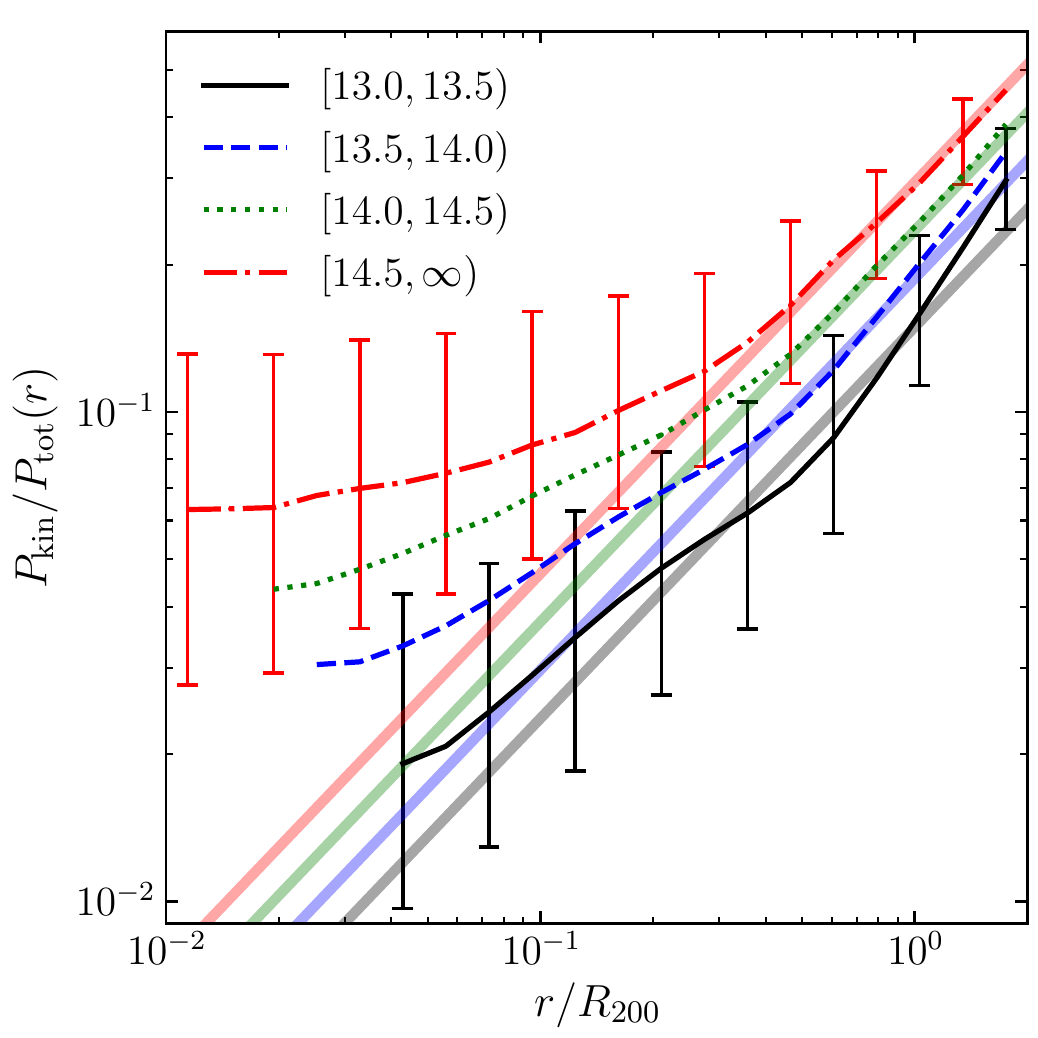}
\end{center}
\caption{Fraction of the kinetic pressure with respect to the total
(kinetic plus thermal) pressure for each of the four mass bins in the 
NR run at $z = 0$. The lighted shaded solid lines trace the fitting
function from \citet{battaglia/etal:2012a} which is based on power-law 
slopes of $4/5$ in radius and $1/5$ in mass. The fitting function is 
computed using the median halo mass in each bin. We show the 25th
and 75th percentiles of the highest (lowest) mass bin as the red 
(black) error bars.}
\label{fig:pkinetic}
\end{figure}

We can obtain a better understanding of the HSE mass bias by measuring the
non-thermal pressure for each individual halo. 
Following \citet{battaglia/etal:2012a}, we focus here on
bulk flows which should capture the major contribution to non-thermal
support. We compute the corresponding kinetic pressure based on 
mass-averaged velocity fluctuations:
\bq
P_{\rm kin} = \frac{\rho_b}{3} 
\left\langle \delta v \cdot \delta v \right\rangle
\label{eq:pkinetic}
\eq
The velocity fluctuations are made with respect to the baryon center of
mass, $\bar{x}$, and mass-averaged velocity, $\bar{v}$, within $R_{200}$:
\bq
\delta v = a (v - \bar{v}) + \dot{a}(x - \bar{x}).
\label{eq:deltav}
\eq 
The resulting kinetic pressure profiles are shown in Figure \ref{fig:pkinetic}
as the fraction of the total pressure. These are compared to the fitting
function from \citet{battaglia/etal:2012a} which is based on the
\citet{shaw/etal:2010} power-law fitting function with an additional mass
dependency that scales as $M_{200}^{1/5}$. 

The fitting function does a good job at matching our simulation
results for $r/R_{200} \gtrsim 0.4$, but dramatically underestimates the kinetic
pressure on smaller scales. In this case, a broken power-law with a shallower slope
on small scales would be more appropriate. We expect the NR results to 
be accurate for $r/R_{200} \gtrsim 0.1$ since bulk flows on these 
scales will be gravitationally sourced and mostly unaffected by 
cooling and feedback processes occurring on smaller scales 
\citep{battaglia/etal:2012a,nelson/etal:2014}. Indeed, the fitting function
in Figure \ref{fig:pkinetic} was calibrated against simulations
with cooling and feedback. Earlier works have shown that kinetic pressure 
is greater in less relaxed systems which have undergone recent major mergers
\citep{jeltema/etal:2008,piffaretti/valdarnini:2008,lau/etal:2009,nelson/etal:2014}.
The rise of the kinetic pressure fraction with
increasing mass is consistent with the picture that high-mass systems are more
likely to be unrelaxed and dynamically disturbed. This is also consistent with
the previous finding that the mass bias is higher for more massive halos.
Moreover, the large error bars demonstrate that a considerable amount of 
halo-to-halo scatter exists in the kinetic pressure fraction.


\section{Sunyaev-Zel'dovich Effect}
\label{sec:sz}

\begin{figure*}
\centering
\includegraphics[width=\smhwidth]{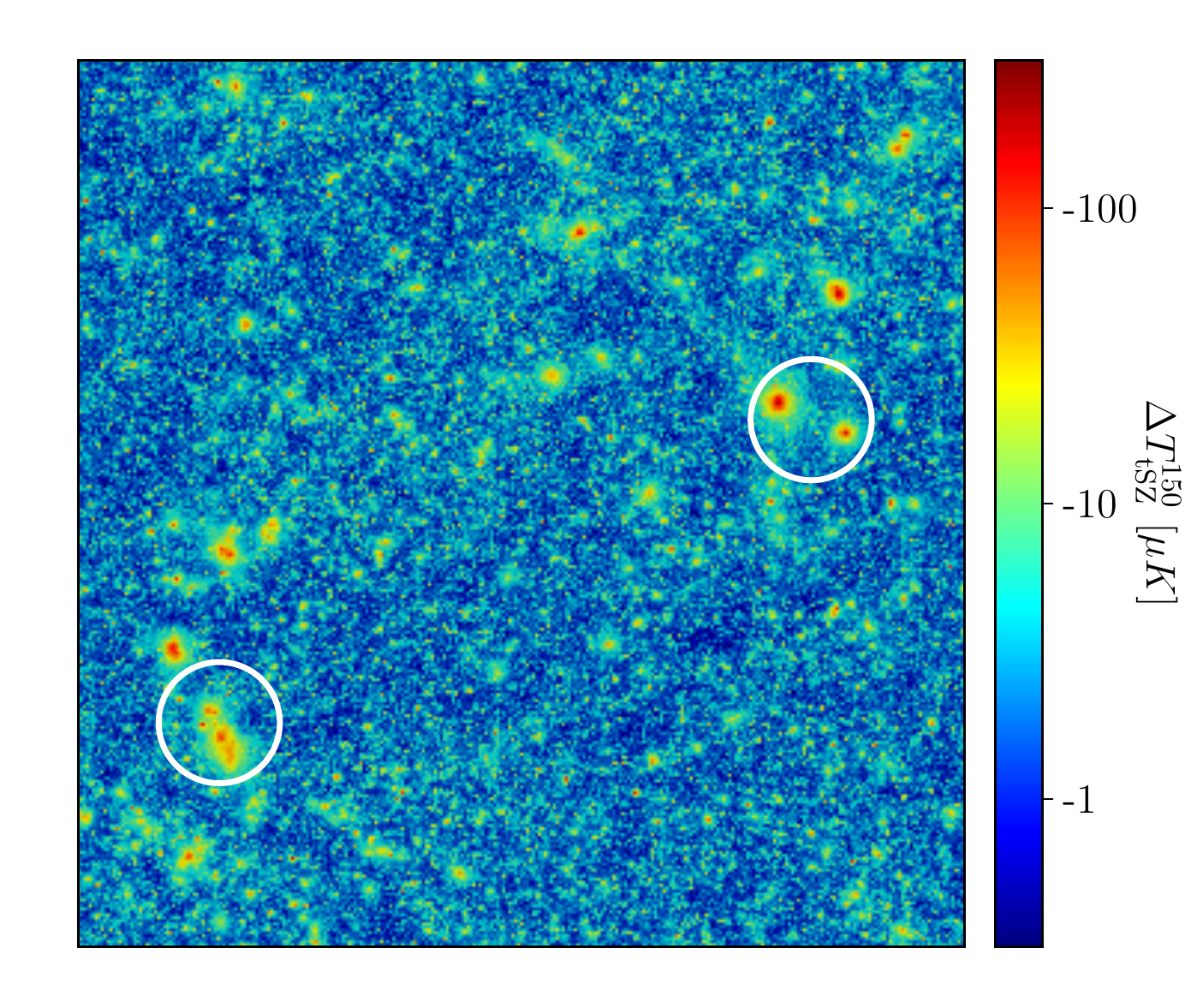}
\includegraphics[width=\smhwidth]{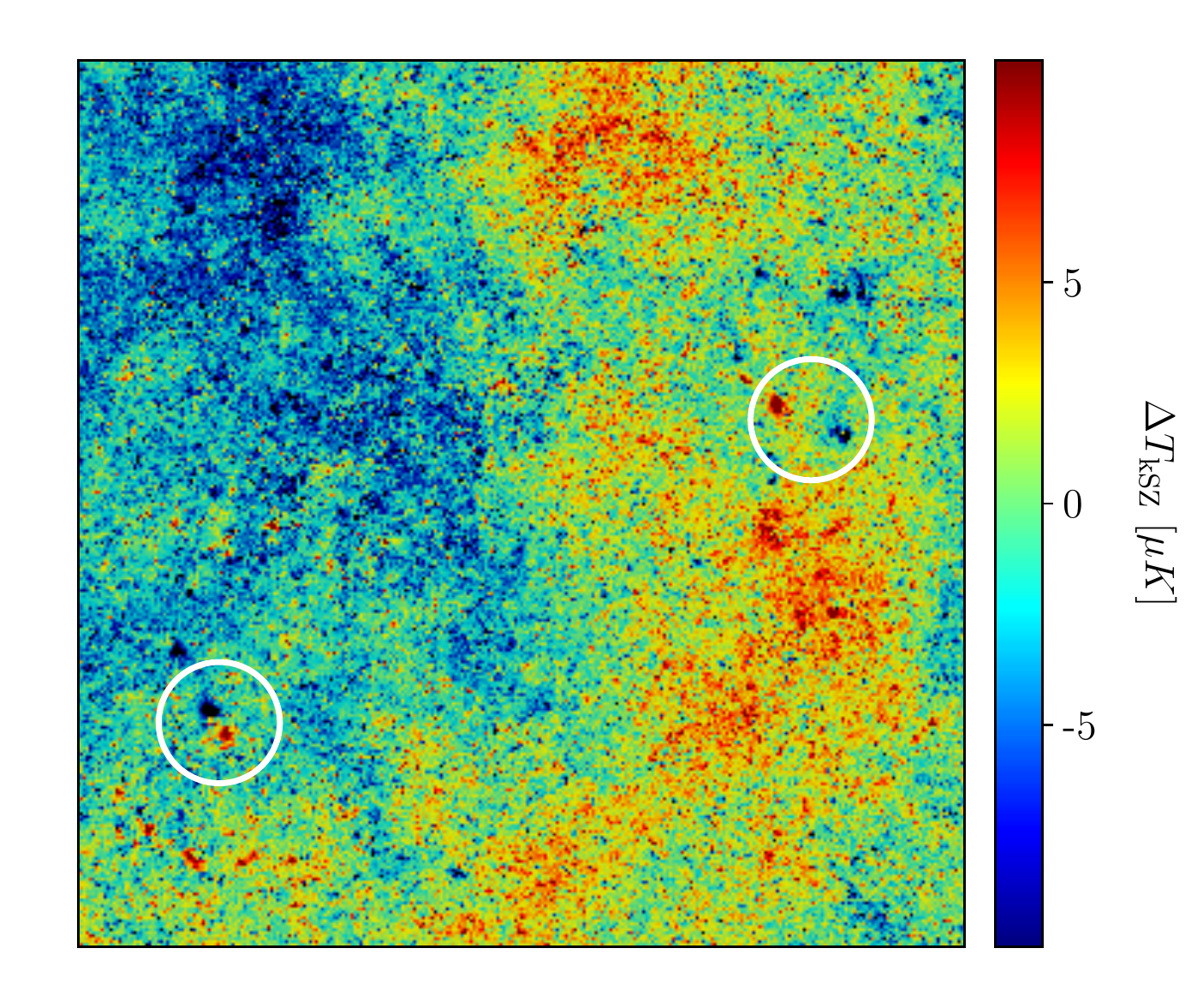}
\caption{tSZ (left) and kSZ (right) temperature fluctuations for a
7.3$^\circ$$\times$7.3$^\circ$ patch of the sky obtained by integrating through a particle
lightcone from the NR Borg Cube run. The tSZ is shown at an observing
frequency of $\unit{150}{\GHz}$ where the signal is a decrement with
respect to the CMB. Note the clear
visibility of clusters in the tSZ map which correspond to hot and cold spots
in the kSZ map. Also visible is the pairwise kSZ signal associated with nearby
clusters. White circles in each image denote the location of two nearby clusters 
whose pairwise signal appears in the kSZ map.}
\label{fig:szmaps}
\end{figure*}

CMB photons interact with matter during their passage between the surface of 
last scattering and today, resulting in a set of secondary anisotropies that 
overwhelm the primary CMB signal on small angular scales. The main contributor
is the Sunyaev-Zel'dovich (SZ) effect which is usually separated into its
thermal (tSZ) and kinematic (kSZ) components
\citep{sunyaev/zeldovich:1970,sunyaev/zeldovich:1972}. The SZ effect is sourced
by inverse Compton scattering off free electrons with the thermal component
weighted towards hot cluster gas whereas the kinematic component is more sensitive
to bulk flows. Hence, together the integrated tSZ and kSZ signals contain a wealth 
of information regarding the evolution of structure formation and the ionization 
state of the universe. 

The magnitude of the thermal and kinematic components can be described in terms of the 
dimensionless Compton $y$ and Doppler $b$ parameters, respectively. The former 
involves an integration of electron pressure along the line-of-sight ($\hat{n}$):
\bq
y = \frac{\sigma_T}{m_e c^2} \int P_e(l) dl
= \frac{\sigma_T k_B}{m_e c^2} \int n_e(l) T_e(l) dl,
\label{eq:tsz}
\eq
while the latter involves an integration over the electron peculiar velocity:
\bq
b = -\frac{\sigma_T}{c}\int n_e (v_e\cdot\hat{n}) dl.
\label{eq:ksz}
\eq
Here $\sigma_T$ is the Thomson cross section, $k_B$ the Boltzmann constant, $m_e$ 
the electron mass, $c$ the speed of light, $n_e$ the free electron number density,
$T_e$ the electron temperature, and $(v_e\cdot\hat{n})$ the electron peculiar
velocity projected along the line-of-sight. We use the convention that $v_e > 0$
for gas moving away from the observer. The electron number density is related
to the gas density, $\rho_g$, via:
\bq
n_e = \frac{\chi \rho_g}{\mu_e m_p},
\label{eq:ne}
\eq 
where $m_p$ is the proton mass and $\mu_e = 1.14$. $\chi$ is the fraction of
electrons that are ionized and is derived from the primordial helium abundance, $Y$,
and the number of electrons ionized per helium atom, $N_{\rm He}$, as:
\bq
\chi = \frac{1-Y(1-N_{\rm He}/4)}{1-Y/2}.
\label{eq:chie}
\eq 
This expression assumes that hydrogen is fully ionized. In what follows, we take
$Y = 0.24$ and assume that helium is neutral ($N_{\rm He} = 0$) so that $\chi = 0.86$.
We also implicitly work in the non-relativistic regime. 

The SZ temperature fluctuations about the CMB monopole are recovered from 
$y$ and $b$ as:
\bq
\frac{\Delta T}{T_{\rm CMB}}(\hat{n}) = 
\frac{\Delta T_{\rm tSZ}}{T_{\rm CMB}} +
\frac{\Delta T_{\rm kSZ}}{T_{\rm CMB}}
= f_\nu y + b.
\label{eq:dtcmb}
\eq 
The function $f_\nu$ encapsulates the frequency dependence of the tSZ signal:
\bqa 
f_\nu &\equiv& x_\nu {\rm coth}(x_\nu/2) - 4, \nn
x_\nu &\equiv& h\nu/(k_B T_{\rm CMB}),
\label{eq:fnu}
\eqa 
where $\nu$ is the observing frequency and $h$ is the Planck constant. The tSZ spectral 
dependence has a null at $\nu_0 \approx \unit{218}{\GHz}$ with temperature
decrements (increments) occurring at frequencies less (greater) than $\nu_0$. 
The kSZ signal, in contrast, is independent of frequency for the non-relativistic 
case assumed here.

\begin{figure*}
\centering
\includegraphics[width=\smfwidth]{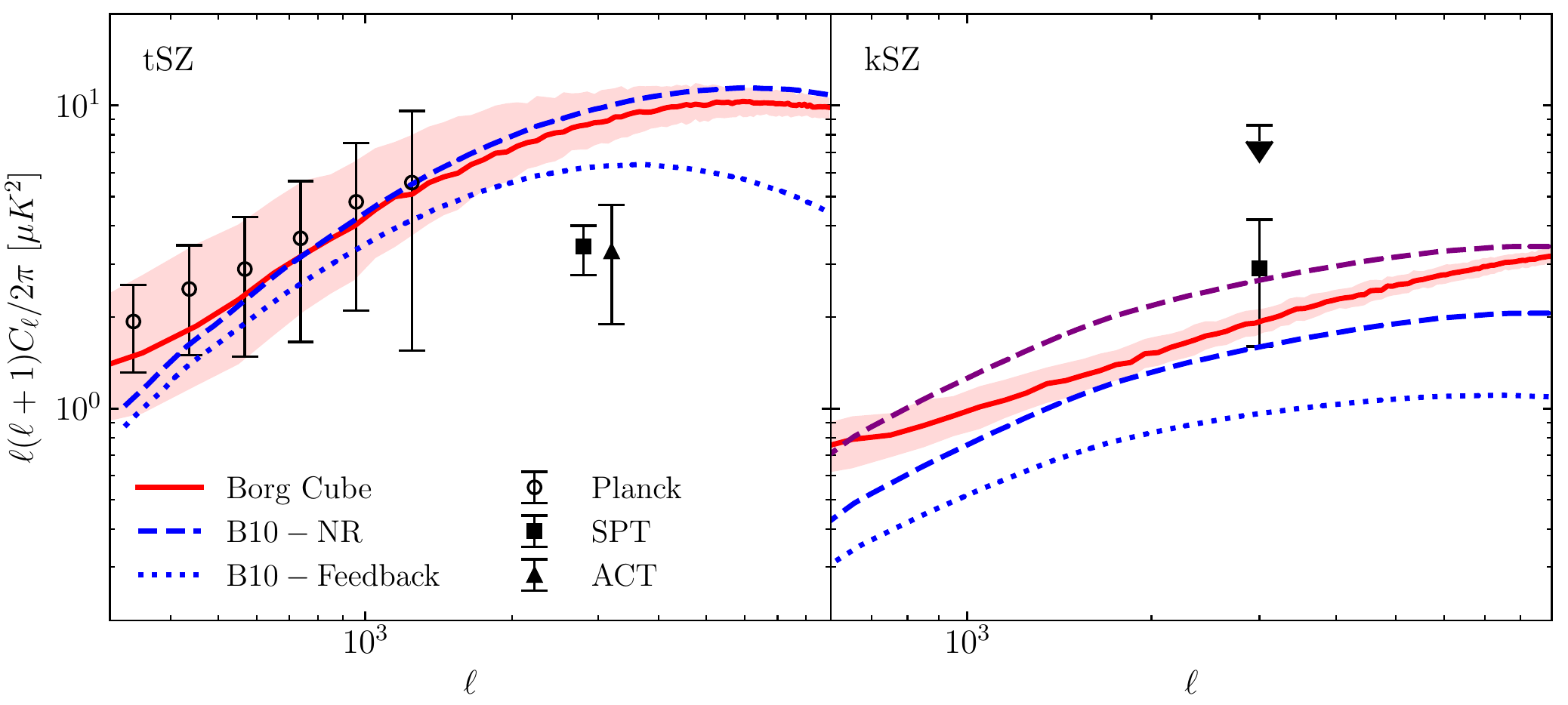}
\caption{Angular power spectra for tSZ at $\unit{150}{\GHz}$ (left) and kSZ 
(right). The solid red lines trace the median result for power spectra computed in 83
independent maps of equal area $\unit{53.7}{\sqdeg}$ while the shaded region
shows the $1\sigma$ spread. The dashed (dotted) blue line traces the NR (feedback)
simulation of \citet{battaglia/etal:2010} which have been appropriately scaled
to the Borg Cube cosmology (see text). Circles with error bars show the low-$\ell$ \planck\ 
tSZ measurements \citep{planck/etal:2016} while squares and triangles show the $\ell = 3000$ 
constraints from SPT \citep{george/etal:2015} and ACT \citep{dunkley/etal:2013},
respectively. The SPT and ACT tSZ constraints have been shifted horizontally
for clarity. The ACT kSZ constraint corresponds to an upper limit.
The dashed purple line in the right panel shows the \citet{battaglia/etal:2010}
NR kSZ result rescaled by the \citet{shaw/etal:2012} correction of 
$67\%$ estimated to account for finite volume effects.}
\label{fig:dlspectra}
\end{figure*}

We generate synthetic $y$ and $b$ maps by integrating equations (\ref{eq:tsz})
and (\ref{eq:ksz}) through a particle lightcone covering one octant of the sky.
In order to fill the entire volume, we stack the simulation box while applying
random rotations to each replicant so as to avoid repeating the same structure along 
the line-of-sight. We compute the temperature of each baryon particle at the time of
lightcone crossing by linearly interpolating its value at the two snapshots
adjacent to the crossing. Velocities are computed from the difference in
particle position at the two adjacent snapshots. We evaluate
equation (\ref{eq:tsz}) from redshift $z = 0.1$ to 5 using a total of 71 particle
snapshots while equation (\ref{eq:ksz}) is evaluated from $z = 0.1$ to 3 using 54
snapshots. In each case, the lower bound is chosen to remove large variance
associated with the possibility of a massive cluster appearing in the field of view
at low redshift. The $b$ map is integrated up to only $z = 3$ since we found that the
finite box size creates visually large velocity discontinuities along the boundaries 
of the stacked boxes. This issue becomes worse at higher redshift as the angle
subtended by the box decreases; the angular extent of the box at $z = 3$ is roughly
$10^\circ$. This issue is much less apparent in the $y$ map since the 
density and temperature fields are relatively smooth on large scales, allowing us
to integrate to higher redshift. The integrated $y$ and $b$ maps are then projected
onto the sky using a \healpix\ \citep{gorski/etal:2005} grid with $\unit{0.184}{\sqamin}$
resolution ($N_{\rm side} = 8192$). The lightcone is generated using all
baryon particles for $z \leq 1$ while sampling at a rate of $50\%$ for 
$1 < z \leq 2$, $25\%$ for $2 < z \leq 3$, and $12.5\%$ for $z > 3$. 

Figure \ref{fig:szmaps} shows $7.3^\circ$$\times$$7.3^\circ$ Cartesian projections
of tSZ and kSZ temperature fluctuations from the NR run. The tSZ map is shown at
frequency $\nu = \unit{150}{\GHz}$ for consistency with SPT and ACT observations;
at this frequency, temperature fluctuations are decrements with
respect to the CMB. Since the tSZ signal is weighted toward hot intracluster gas, 
it is easy to visually pick out the strong temperature decrements associated with
the cores of massive clusters. The location of these clusters
also appear as hot and cold spots in the kSZ map. In addition, it is possible to see
the presence of the pairwise kSZ signal associated with nearby clusters whose velocity 
vectors point in opposite directions along the line-of-sight, creating paired
hot and cold spots \citep{flender/etal:2016}. Note that this analysis involves only the
post-reionization kSZ signal and thus ignores important contributions from the
``patchy'' network of ionized bubbles around luminous sources during the epoch
of reionization. In general, the magnitude and shape of the patchy kSZ signal 
will depend on the duration of reionization and the size distribution of ionized 
bubbles \citep{mcquinn/etal:2005,zahn/etal:2005,iliev/etal:2007}.

We provide a more quantitative analysis by plotting the tSZ and kSZ angular
power spectra in Figure \ref{fig:dlspectra}. Power spectra are computed in 83
independent maps of equal area $\unit{53.7}{\sqdeg}$ with the 
solid red line showing the median of these maps and the shaded region
showing the $1\sigma$ scatter. For comparison, the dashed (dotted) blue line
shows the NR (feedback) result from the simulations of
\citet[][hereafter B10]{battaglia/etal:2010}. 
Circles with error bars show the low-$\ell$ \planck\ measurements 
\citep{planck/etal:2016} while the squares and triangles show the
$\ell = 3000$ constraints from SPT \citep{george/etal:2015} and 
ACT \citep{dunkley/etal:2013}, respectively. In all cases, the tSZ power has been
appropriately adjusted to $\unit{150}{\GHz}$ using equation (\ref{eq:fnu}).

We find good agreement between the low-$\ell$ \planck\ tSZ
measurements and the Borg Cube run. In contrast, our $\ell = 3000$ power
is considerably higher than the SPT and ACT constraints. This is 
expected given that our NR simulation does not include feedback which reduces
gas pressure in cluster cores. This can be seen by comparing the NR and feedback
simulations of B10. The agreement between the Borg Cube and B10 NR simulation 
is encouraging. There does appear to be a trend toward slightly smaller power 
at the highest $\ell$ in our run, which may reflect differences in the hydro solver
(B10 use the tSPH scheme of \gadget). Of course, these scales will be highly
sensitive to feedback processes. It is important to control for cosmology when 
comparing to other works since the tSZ effect is strongly cosmology dependent. 
We use the same $\sigma_8$ as B10, but have minor differences in baryon density, 
which we account for here by scaling the B10 result by 
$C_\ell \propto (\Omega_b h)^2$ \citep{komatsu/seljak:2002}.

The kSZ signal has a significantly lower amplitude than its thermal counterpart. 
In our case, we find the $\ell = 3000$ kSZ to be $22\%$ that of the
$\unit{150}{\GHz}$ tSZ signal. We again compare our result to the B10 NR run and
adjust for cosmology using the 
$C_\ell \propto \Omega_b^{2.13}h^{1.68}z_r^{0.43}$ 
scaling relations suggested in \citet{shaw/etal:2012}. Here $z_r$ corresponds
to the upper redshift limit which we take as $z_r = 3$ while B10 use $z_r = 10$.
We find reasonable agreement between the Borg Cube and B10 NR curves though our
result is systematically higher on all scales. One possibility for this discrepancy
may be related to our larger simulation volume. \citet{shaw/etal:2012} show that the 
truncation of large-scale velocity modes by finite 
simulation boxes can drastically underestimate the kSZ amplitude. They estimate a $67\%$ 
enhancement of kSZ power at $\ell = 3000$ to compensate for the $\unit{165}{\Mpch}$ box 
used in B10. In our case, an enhancement closer to $\approx5\%$ would be more appropriate. 
The dashed purple line in Figure \ref{fig:dlspectra} shows the B10 NR result rescaled by 
a constant $67\%$. It is clear that finite volume effects have the potential to induce
large changes on the simulated kSZ signal.
Without comparing our result to a larger box, it is difficult to determine
how much the discrepancies between our curves are driven by finite volume effects versus 
choices in the kSZ integration scheme or differences in the hydro solvers.

Our kSZ result at $\ell = 3000$ is high compared to the SPT constraint given 
the fact that our analysis is missing power associated with integrating up to only 
$z = 3$  and we additionally ignore contributions from patchy reionization
\footnote{The \citet{shaw/etal:2012} scaling relation suggests
that integrating the non-patchy signal up to e.g. $z = 10$ would increase kSZ power
by $68\%$. Depending on the details of patchy reionization, the signal can further
increase by another factor of $\sim2$ \citep[e.g.][]{iliev/etal:2007}.}.
However, this high result would be partly offset by power suppression associated with 
feedback (as seen by comparing the B10 curves). Also recall that our analysis ignores 
power from free electrons associated with helium reionization which in principle would 
shift both our tSZ and kSZ curves up by a constant factor. 


\section{Summary}
\label{sec:summary}

Constraining the nature of dark energy requires an accurate understanding
of the impact of baryons on cosmological structure formation. This is a
nontrivial task due to the high dynamic range of spatio-temporal scales involved
and the complexities of the underlying astrophysics. Cosmological
hydrodynamic simulations are the best option for modeling baryonic 
processes with high mass resolution in representative volumes. The simplest
treatment involves NR physics where the thermal state of baryons changes
only in response to gravitational shocks and the expansion of the universe. Even in this
case, uncertainties arise due to systematics in the hydro solver.
It is therefore crucial to compare the results from different codes in order to 
gauge the accuracy of predictions derived from simulations. This may also involve
using the same code, but with different implementations (e.g., hydro solver, baryonic physics) 
so that changes between each comparison are minimized and systematics more easily 
isolated. The potential for systematics increases with the inclusion
of subresolution physics treatments
since a large degree of freedom exists in these models. 

In this paper, we have presented results from the Borg Cube run
which is the first cosmological hydrodynamic simulation based on the
CRK-SPH formalism. We have restricted attention to the NR case since this
provides the most parameter-free comparison point for hydro solvers. 
We have studied various statistics of the evolved matter field and drawn
comparisons to previous simulations where available. We also compared the NR results to
a GO version of the Borg Cube which used identical initial conditions.
This is useful not only in showing the relative impact of shock heating, 
but also in identifying possible systematics that may arise from multi-species
interactions. We report the main conclusions from these investigations in the
following paragraphs.

{\em Power spectra}: the NR run shows a strong suppression (modest increase) 
of baryon (CDM) power on small scales relative to the GO case. 
This is in qualitative agreement with the earlier simulations of 
\citet{jing/etal:2006} and \citet{vogelsberger/etal:2014} which were based on 
tSPH and moving-mesh methods, respectively. The moving-mesh and CRK-SPH methods 
agree quite well on the details of baryon suppression at $z = 0$ with 
$\approx10\%$ suppression at $k \sim \unit{1}{\invMpch}$ followed
by a steep drop-off on scales $k \gtrsim \unit{5}{\invMpch}$. In contrast,
the tSPH result displays more modest suppression on the 
smallest scales. This likely owes to the fact that tSPH has been
shown to over-shoot baryon clustering within the cores of collapsed objects
\citep{frenk/etal:1999,sembolini/etal:2016}. The moving-mesh and CRK-SPH
methods match very well in terms of CDM and total matter power with only
few-percent differences emerging on scales $k \gtrsim \unit{2}{\invMpch}$.
Of course, it is precisely on such small scales where contributions from cooling 
and feedback lead to large changes compared to the NR case.

{\em Matter redistribution within halos:} shock heating induces an internal
redistribution of both baryons and CDM within collapsed objects. We have shown
that, relative to the GO case, this redistribution follows a roughly universal
form independent of halo mass. To begin, both components exhibit a 
few-percent dip in density at $R_{200}$ relative to the GO run. For baryons,
this is followed by an inward increase in relative density that peaks at a 
level of $5\%$ at $\approx 0.4R_{200}$ before sharply dropping to a value
about one-tenth that of the GO run in the halo center. The CDM relative density
also rises inward from $R_{200}$ and peaks at a $5\%$ enhancement while 
remaining near this level throughout most of the radial range. Since halos 
are defined with respect to a constant over-density, this redistribution 
leads to a systematic shift in the radius and mass of each SO halo.  
We find an average increase of $1\%$ in $M_{200}$ with a corresponding
few-percent change in the mass function. These numbers agree well with the
tSPH simulations of \citet{cui/etal:2012}. We can also describe these changes
in terms of the concentration. We find that $c_{200}$ is reduced by $2\%$
($7\%$) compared to the GO run for halos of mass 
$\sim\unit{10^{15}~(10^{13})}{\massh}$. This is in contrast to the $5-10\%$ 
{\em increase} in $c_{200}$ found in earlier works 
\citep{rudd/etal:2008,rasia/etal:2013}; a discrepancy that is at least 
partly sourced by differences in hydro solvers.

{\em Baryon fraction:} we find the global baryon fraction within $R_{200}$
at $z = 0$ to be about $95\%$ the universal mean for halos 
$M_{200} \geq \unit{10^{13}}{\massh}$. Previous tSPH simulations find
somewhat smaller values in the range $89-93\%$ 
\citep{ettori/etal:2006,crain/etal:2007,stanek/etal:2010,battaglia/etal:2013}
It has been shown that 
mesh-based methods produce baryon fractions $\approx5\%$ higher than tSPH
\citep{frenk/etal:1999,kravtsov/etal:2005,stanek/etal:2009}; CRK-SPH also seems 
to fall within this camp. Likewise, the recent code comparison of
\citet{sembolini/etal:2016} showed that other modern SPH treatments produce 
baryon fractions more consistent with mesh-based codes.

{\em Self-similar gas profiles:} we find that stacked baryon density, 
temperature, entropy, and pressure profiles show self-similar results
across all of the converged mass bins spanning two orders of magnitude in mass. 
The density profiles tend to cores within $0.1R_{200}$ with relatively 
constant concentrations $c_{200} \approx 3$. Temperature is found to slowly 
rise inward and approach a central value about three times larger than 
that at $R_{200}$. Entropy decreases inward with a power-law slope of 
1.2 in radius before approaching an isentropic core within $0.1R_{200}$. 
Pressure rises strongly with decreasing radius reaching central values 
three orders of magnitude  larger than at $R_{200}$. We attribute this 
self-similarity across mass to the fact that both the underlying NFW 
matter distribution and NR physics are relatively scale-free. This will not be the 
case when cooling and feedback prescriptions are included. 
As such, the NR results shown here are mostly applicable to halo outskirts 
for which $r \gtrsim 0.5R_{200}$. Indeed, \citet{burns/etal:2010} confirm
that NR simulations are sufficient at modeling observed cluster profiles
on these scales.

{\em Non-thermal pressure support:} a simple model based on the assumption 
of HSE is capable of 
predicting density, temperature, entropy, and pressure within
$40\%$ over two orders of magnitude in halo mass. In general, the model does
not perform well around $R_{200}$ due to the breakdown of the assumption of 
full thermal pressure support. Similarly, observational estimates of cluster 
mass based on the assumption of HSE will be biased. The main contribution to 
non-thermal pressure comes from turbulent and bulk flows that develop during 
the structure formation process. Measuring this directly in the Borg Cube, 
we find $20-40\%$ of the total pressure support at $R_{200}$ comes from 
kinetic pressure. On average, this is higher for more massive halos with a 
power-law scaling of $1/5$ in $M_{200}$ as suggested in the feedback 
simulation of \citet{battaglia/etal:2012a}. The fraction of 
kinetic pressure drops with decreasing radius. We find HSE mass estimates are 
biased low by $24\%$ ($10\%$) for halos of mass 
$\sim\unit{10^{15}~(10^{13})}{\massh}$. This is in
agreement with previous NR simulations based on both tSPH and AMR methods
\citep{kay/etal:2004,nelson/etal:2012}. The average bias does not evolve 
strongly within the radial range $[0.1-1]R_{200}$ though a considerable amount
of halo-to-halo scatter ($10-15\%$) exists.

{\em Sunyaev-Zel'dovich Effect:} we compute angular power spectra for both
the tSZ and kSZ effects by integrating through a particle lightcone from the
Borg Cube run. Our NR results are most applicable to multipoles $\ell \lesssim 1000$
where we find good agreement with \planck\ tSZ measurements. Our tSZ result
also shows excellent agreement with the NR simulation of B10 based on tSPH. This
suggests that the tSZ signal derived from simulations is relatively insensitive to
the choice of hydro solver. We find relatively good agreement with the B10 NR kSZ
result though the comparison is difficult in this case since the small box of B10
is heavily impacted by artificial suppression of large-scale velocity modes.
A comparison to a box of similar size as the Borg Cube would be needed to address
how sensitive the simulated kSZ signal is to the choice of hydro solver. 
The Borg Cube SZ predictions at $\ell = 3000$ are high compared to
SPT and ACT constraints, which is expected given our omission of feedback.

{\em Artificial particle coupling:} special care must be taken to control
numerical artifacts that may occur between particle species with unlike mass
and/or initial power spectra. In both our GO and NR simulations, we employed
a common approach of using a constant gravitational softening length for all
particle pairs. It is easy to test for artifacts in the GO run since the CDM 
and baryons are effectively equivalent in this case. Artificial 
scattering between the heavy CDM and light baryons leads to an increase (decrease)
in clustering for the former (latter) species. This is evident in individual 
power spectra as well as radial profiles of the baryon fraction.
This issue propagated out to scales of about seven times the gravitational softening 
length in the GO run. The total matter distribution, on the other hand, was
converged down to smaller scales of about twice the softening length suggesting that
this process operates in such a way as to mostly preserve the total matter field.
Furthermore, halos less massive 
than 3200 times the combined CDM plus baryon particle mass failed to converge 
in terms of the global baryon fraction. Isolating this effect is more difficult 
in the NR run and we did not see any clear systematics in radial profiles
of the baryon fraction. It is plausible that the SPH smoothing kernel and
artificial viscosity somewhat regulate the issue. We leave a more thorough
investigation into the interplay between gravitational and hydrodynamic 
interactions to future work.

This work focused on the NR case and is thus valid on
a limited range of scales. Including all the physics relevant to galaxy 
formation is a formidable challenge and one for which progress must be made in
a controlled manner. On cosmological scales, the only path forward is through
sub-resolution treatments of cooling and feedback. In this case, a large degree
of modeling freedom exists and the potential for being systematics-limited increases. We 
use this work as the first stepping stone toward more sophisticated gas treatments 
within the \crkhacc\ formalism.


\acknowledgments{
We thank Matt Becker, Lindsey Bleem, Jon\'{a}s Chaves-Montero, Hillary Child, Andrew Hearin,
Joe Hollowed, and Steve Rangel for helpful discussions. We are indebted to Volker Springel for providing the {\arepo}  comparison data set. SH acknowledges inspiring past discussions with Bryan (Bucky) Kashiwa in Los Alamos National Laboratory's Theoretical Division.
Work at Argonne National Laboratory was supported under U.S. Department of Energy
contract DE-AC02-06CH11357.
An award of computer time was provided by the Theta Early Science Program (ESP).
This research used resources of the Argonne Leadership Computing Facility, which is 
a DOE Office of Science User Facility supported under Contract DE-AC02-06CH11357. 
Part of this research was supported by the Exascale Computing Project (17-SC-20-SC), 
a collaborative effort of the U.S. Department of Energy Office of Science and the
National Nuclear Security Administration.
This work is part of the Borg Collective. 
}


 \newcommand{\noop}[1]{}

\end{document}